\pdfoutput=1 
\documentclass[a4paper,11pt]{article}

\usepackage[utf8]{inputenc}

\usepackage{jcappub} 
\usepackage{threeparttable}
\usepackage{graphicx,rotating,wrapfig,float}
\usepackage{tabularx}
\usepackage{multirow}
\usepackage{cleveref}
\usepackage{units}
 \usepackage{subfigure}

\title{The \ardm\ Liquid Argon Time Projection Chamber at the Canfranc Underground Laboratory: a ton-scale detector for Dark Matter Searches}

\collaboration{The ArDM Collaboration:}

\author[a]{J.~Calvo}
\author[a]{C.~Cantini}
\author[a]{P.~Crivelli}
\author[b]{M.~Daniel}
\author[a]{S.~Di~Luise}
\author[a]{A.~Gendotti}
\author[a]{S.~Horikawa}
\author[b]{B.~Montes}
\author[a]{W.~Mu}
\author[a]{S.~Murphy}
\author[a]{G.~Natterer}
\author[a]{K.~Nguyen}
\author[a]{L.~Periale}
\author[a]{Y.~Quan}
\author[a]{B.~Radics}
\author[a]{C.~Regenfus}
\author[b]{L.~Romero}
\author[a]{A.~Rubbia}
\author[b]{R.~Santorelli}
\author[a]{F.~Sergiampietri}
\author[a]{T.~Viant}
\author[a]{S.~Wu}

\affiliation[a]{ETH Zurich, Institute for Particle Physics, Zurich, Switzerland}
\affiliation[b]{CIEMAT, Div. de F{\'\i}sica de Particulas, Avda. Complutense, 22, E-28040, Madrid, Spain}

\emailAdd{andre.rubbia@cern.ch}

\abstract{
The Argon Dark Matter (\ardm) experiment consists of a liquid argon (LAr) time projection chamber (TPC) sensitive to nuclear recoils resulting from scattering of hypothetical Weakly Interacting Massive Particles (WIMPs) on argon targets. With an active target of 850\,kg, \ardm\ represents an important milestone in the quest for Dark Matter with LAr. We present the experimental apparatus currently installed underground at the \textit{Laboratorio Subterr\'aneo de Canfranc} (LSC), Spain. We show first data recorded during a single-phase commissioning run in 2015 (\ardm\ \rI), which overall confirm the good and stable performance of the ton-scale LAr detector. 
}

\keywords{Dark Matter; WIMP; nuclear recoils; LAr TPC; ArDM}

\newcommand{\alp}{\ensuremath{\alpha}}
\newcommand{\bet}{\ensuremath{\beta}}
\newcommand{\gam}{\ensuremath{\gamma}}
\newcommand{\ardm}{ArDM}
\newcommand{\nier}{NIER}

\newcommand{\sala}{\textit{Sala A}}
\newcommand{\rI}{Run\,I}
\newcommand{\kr}{\ensuremath{\,\mathrm{^{83m}Kr}}}
\newcommand{\ar}{\ensuremath{\,\mathrm{^{39}Ar}}}
\newcommand{\co}{\ensuremath{\,\mathrm{^{57}Co}}}

\newcommand{\kev}{\ensuremath{\,\mathrm{keV}}}
\newcommand{\mev}{\ensuremath{\,\mathrm{MeV}}}
\newcommand{\gev}{\ensuremath{\,\mathrm{GeV}}}
\newcommand{\cm}{\ensuremath{\,\mathrm{cm}}}
\newcommand{\mum}{\ensuremath{\,\mu\mathrm{m}}}

\newcommand{\mgcs}{\ensuremath{\,\mathrm{mg/cm^{2}}}}
\newcommand{\mm}{\ensuremath{\,\mathrm{mm}}}
\newcommand{\nm}{\ensuremath{\,\mathrm{nm}}}
\newcommand{\mis}{\ensuremath{\,\mathrm{ms}}}
\newcommand{\mus}{\ensuremath{\,\mu\mathrm{s}}}
\newcommand{\ns}{\ensuremath{\,\mathrm{ns}}}
\newcommand{\kv}{\ensuremath{\,\mathrm{kV}}}
\newcommand{\sig}{\ensuremath{\,\sigma}}
\newcommand{\ohm}{\ensuremath{\,\Omega}}
\newcommand{\hz}{\ensuremath{\,\mathrm{Hz}}}
\newcommand{\khz}{\ensuremath{\,\mathrm{kHz}}}
\newcommand{\mhz}{\ensuremath{\,\mathrm{MHz}}}

\newcommand{\trt}{\ensuremath{^{\mbox{\tiny{\textregistered}}} }}

\graphicspath{{./Pictures/}}

\begin{document}


\maketitle

\section{Introduction}
\label{sect:Intro}

The existence of non-luminous, non-baryonic cold Dark Matter is by now well established \cite{Agashe:2014kda}. Several experimental and theoretical indications favour Dark Matter which is supposed to be present in our Galaxy as a halo of the thermal relic from the Big Bang. A popular hypothesis, explaining these observations, is, that Dark Matter is made of Weakly Interacting Massive Particles (WIMPs). However, no such particles exist in the Standard Model and none has been directly observed at particle accelerators or elsewhere\,\cite{Bertone2005279}. Hence the particle physics nature of the WIMPs Dark Matter remains unknown.

A great variety of experiments for direct Dark Matter searches have been running in the recent years utilising different technologies (see e.g.\,\cite{Undagoitia:2015gya} for an overview). The elastic scattering of WIMPs with masses of (10-1000)\gev/$c^2$ is supposed to produce nuclear recoils in the range of (1-100)\kev\,\cite{Lewin199687}. Despite the large experimental effort to search for the rare nuclear recoils in a terrestrial experiment, a conclusive result is still missing, presumably due to the very low interaction probability.

Present searches extend exposures on detector targets to an order of several thousand kg$\cdot$day, with just a few expected background events in the region of interest. This drives developments to larger target sizes, favouring the deployment of liquid noble gas detector using xenon\,\cite{RevModPhys.82.2053,Alner2007287,PhysRevLett.109.181301,PhysRevLett.112.091303,Abe201378,pandax2014} and/or argon\,\cite{Rubbia:2005ge,BenettiWarp2008,Hime:2011ms,deap2012,Agnes:2014bvk} in a TPC for their scalability, cleanliness, and background discrimination power.  

The \ardm\ experiment is designed for highest sensitivity in the WIMPs mass range above 100\,\gev/$c^{2}$. The detector consists of a vertical TPC using LAr as the WIMP target\,\cite{Rubbia:2005ge,Amsler:2010yp,Marchionni:2010fi,Badertscher:2013ygt} and 24 low radioactivity 8'' Hamamatsu PMTs distributed in two equal arrays for light readout. The \ardm\ detector is able to detect signals produced by elastic scattering with LAr atoms in the active volume, of WIMPs or neutrons producing ``nuclear recoils'', and by background particles like \gam\ or \bet{} producing ``electron recoils''. When the detector works in double phases (liquid and gaseous) mode, both nuclear and electron recoils generate scintillation light ($S1$) and electron-ion pairs. The electrons can be separated from their ions in an electric field and drift upwards the argon surface. After being extracted from the LAr to the gaseous argon (GAr) on top, these electrons are accelerated and the secondary scintillation light ($S2$), which is proportional to the amount of electrons extracted, is produced. Both $S1$ and $S2$, which are vacuum ultraviolet (VUV) light with a wavelength around 127\nm, can be wavelength shifted to visible range and read out by the PMT arrays. The pulse shape of $S1$ signal and the ratio $S1/S2$ are different for nuclear and electron recoils~\cite{Boulay:2004dk}, which are used to reject the \gam\ or \bet\ background events. In addition, the $S1$ signals is used to reconstruct the energy of the event, while $S2$ signals help to reconstruct the 3D position of the event.

In 2015 a series of commissioning runs with gaseous and liquid argon targets (\ardm\ \rI ) were undertaken in single phase mode to explore the functionality and performance of the detector. Since no electric field was applied, only $S1$ signals were collected. In this paper, we describe the design and setup of the \ardm\ detector and present results from gaseous data collected during \ardm\ \rI. Results from data taken in liquid are reported elsewhere~\cite{Calvo:2016nwp,ardminprep}.

\section{Experimental setup}
\label{sect:ExpInst}

\subsection{The underground laboratory LSC}
\label{sect:LSC}

In order to reach low background conditions, the \ardm\ experiment is installed at the underground Laboratorio Subterr\'aneo de Canfranc (LSC)\,\cite{Bettini:2012fu}, located under the Mount Tobazo in the central Spanish Pyrenees. The rock overburden is about 2500\,m water-equivalent. The laboratory is situated between an old, decommissioned, railway tunnel and the newly built Somport road tunnel. Access to the underground laboratory is given through one of the emergency safety links connecting both tunnels at regular distances. The recently refurbished LAB 2400 mainly consists of two experimental halls (A and B), with the dimensions of $15\times40$\,m$^{2}$ and $10\times15$\,m$^{2}$ respectively. Table\,\ref{table:LSC} summarises the main parameters of the laboratory\,\cite{Bettini:2012fu,Bettini:2011zza} relevant for underground physics experiments. 

\begin{table}[htb]
\begin{center}
\vspace{2mm}
\begin{tabularx}{\columnwidth}{p{60mm}@{\extracolsep{\fill}}cc} 
\hline
 Muon flux 	                    & 	2-4 	           &	10$^{-3}$\,m$^{-2}\cdot$s$^{-1}$ 	\\
 Neutron flux 	                & 	$3.47 \pm 0.35$  &	10$^{-2}$\,m$^{-2}\cdot$s$^{-1}$ 	\\
 Rn activity in the air 	      &	  50-80 	         &	Bq$\cdot$m$^{-3}$	                \\
 Total \gam\ flux in Hall\,A  	& 	$1.23 \pm 0.17$  &	cm$^{-2}\cdot$s$^{-1}$ 	\\ 
\hline
\end{tabularx}
\caption{Measured background fluxes at Laboratorio Subterr\'aneo de Canfranc (LSC)\,\cite{Bettini:2012fu}.}
\label{table:LSC}
\end{center}
\end{table}

Besides the main experimental halls the underground site also contains a clean room area equipped with various high-purity p-type coaxial germanium counters (HPGe counters). The semiconductor detectors are embedded in lead and copper shieldings and serve as a material screening facility for the experimental components. 

The \ardm\ detector as well as its cryogenic service installation are situated in a lowering of the concrete floor in Hall\,A at LSC (\sala), which serves as a large containment pool in case of accidental loss of the LAr. A second smaller containment volume is created by thermally isolated panels just below the main detector vessel of \ardm. This volume is connected to a gaseous extraction line into the railway tunnel, which can be used for a removal of argon gas in case of an accident or emptying the target.

The laboratory is also equipped with an emergency electrical power supply (Diesel generator) able to sustain the entire installations over several hours without intervention. This includes the \ardm\ cryocoolers consuming about 30\,kW. 

\subsection{Overview of the \ardm\ detector}
\label{sect:Overview}

\begin{figure}[htbp]
\begin{center}
\includegraphics[width=0.9\textwidth]{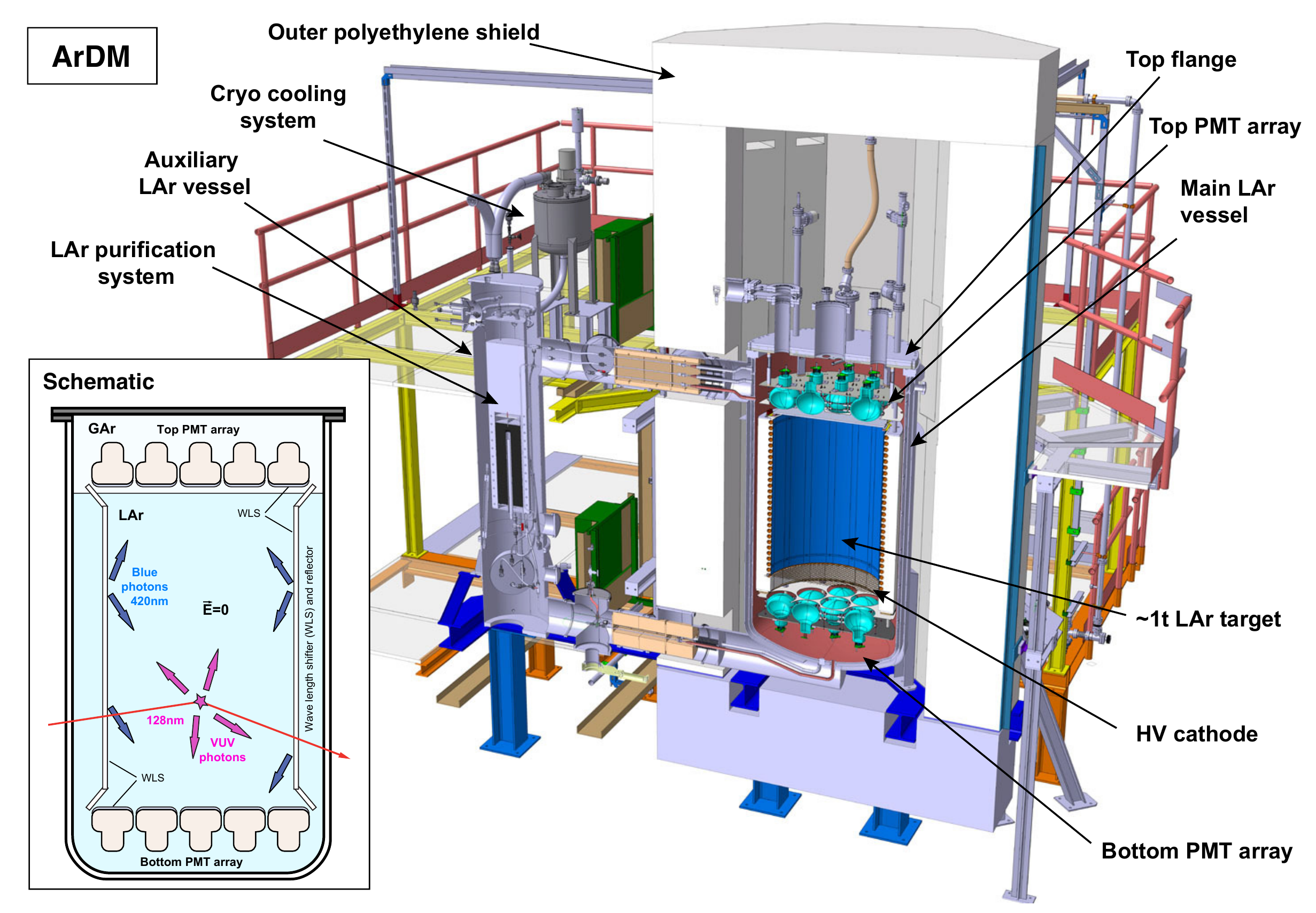}
\caption{Overview of the \ardm\ experimental setup at LSC. The inset to the left shows the schematic of the inner detector in single phase configuration.}
\label{fig:det_structure}
\end{center}
\end{figure}

The main component of the \ardm\ experiment (see Figure\ref{fig:det_structure}) consists of a cylindrical TPC installed in a LAr dewar of 1\,m diameter. A layer of 10\cm\ of LAr is available around the target to shield particles entering from the outside. The detector active volume is confined by an optical surface made of high-reflectivity Polytetrafluoroethylene (PTFE) foils to collect as many photons as possible. The PTFE reflectors are coated with a thin layer of a wavelength shifter (WLS), to convert the argon scintillation VUV light to a range of maximal sensitivity of the PMTs (see Section\,\ref{sect:LRO} for details).

The active target volume, defined by the drift cage, amounts to about 540\,liters, corresponding to about 750\,kg of LAr. For double-phase operation an approximately uniform vertical electric field is created in the active volume. By applying negative HV to the cathode electrons are drifted to the top where they are extracted into the gaseous phase of the detector producing the secondary signal $S2$. The drift cage has a shape of vertical cylinder, 112\cm\ in height and 80\cm\ in diameter, owning a flat section on the side to accommodate the large HV feedthrough. The drift cage is formed by 27 field shaper rings vertically arranged with a pitch of 40\mm . The rings are mounted onto seven 40\mm\ thick pillars made out of high-density polyethylene (HDPE). Top and bottom of the active volume are electrically closed by an extraction and cathode grid, respectively. The maximal design value of the cathode voltage is -100\kv\ creating a drift field up to about 1\kv/cm. A further grid is mounted 13\cm\ below the cathode grid biased to a voltage similar to the one for the PMTs as HV protection.

During \ardm\ \rI\ no voltages were applied to the drift cage ($E$=0) and the detector was operated in single-phase mode with a slightly different geometry than for double phase operation creating an active LAr target of around 850\,kg. 

\subsection{Cryogenic system}
\label{sect:CryoDesign}

\begin{figure}[htb]
\centering
\includegraphics[width=1.0\textwidth]{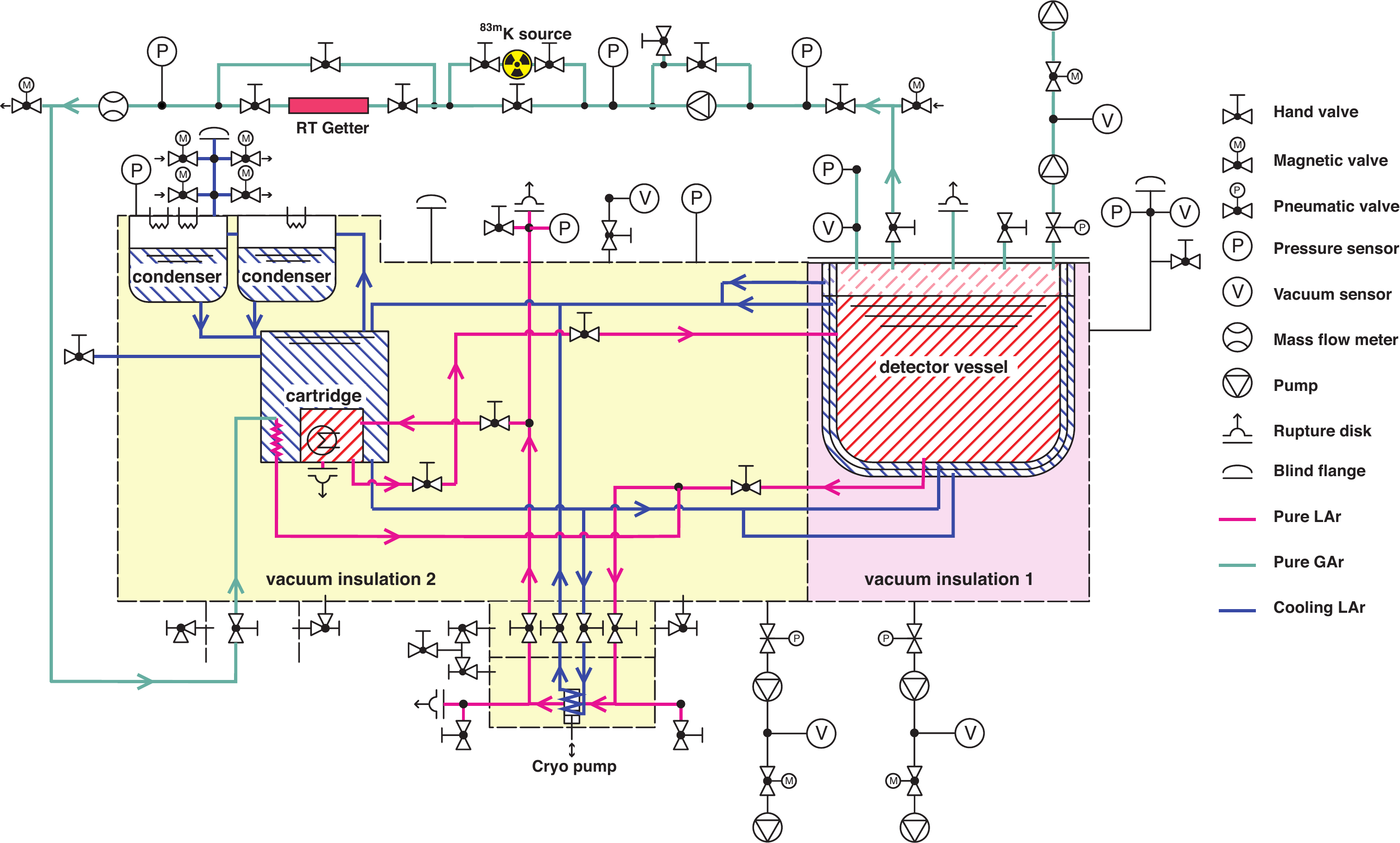}
\caption{Cryogenic, vacuum and instrumentation diagram (P\&ID).}
\label{fig:CryoDesign}
\end{figure}

The \ardm\ high purity cryogenic LAr target is placed in a triple-wall dewar vessel with a LAr layer for cooling\,\cite{Epprecht2012diss}. This bath design, with a separation of clean and dirty LAr volumes is developed to shield direct heat input from the outside with the aim not to create any gas bubbles in the LAr target. Figure\,\ref{fig:CryoDesign} shows the overall schematics of the cryogenic installation of \ardm\ with the main LAr volume on the right and the cryogenic services on the left. Both parts are insulated with separated vacua. The two hermetically closed circuits containing the high purity detector argon (red), as well as the LAr bath used for cooling (blue), are protected against overpressure by electrical and mechanical valves, as well as by rupture disks. 

The LAr bath can be cooled at the same time by three cryocoolers for faster cool-down, however only two are needed for normal operation, leaving the third as spare for safety or during maintenance of a cryocooler. The location of their three cold heads are in the top of the condenser vessels (upper left in Figure\,\ref{fig:CryoDesign}). 

\subsubsection{Detector vessel}
\label{sect:Vessel}

The detector vessel is a three-wall dewar cryostat for LAr, made of stainless steel. The inner main volume, a vertical cylinder of 100\cm\ in diameter and $\sim$200\cm\ in height, can hold up to 1.4\,m$^3$ of LAr. The target space is surrounded by an intermediate LAr layer, the cooling bath, having a thickness of 2\cm. The outer volume serves as insulation vacuum, surrounding the cooling bath. While the bath and the insulation vacuum are closed by welding, the main volume is closed hermetically with a 4-cm-thick stainless-steel top flange with indium seal. The top flange supports the detector structure hanging from it, and has various service ports with feedthroughs all based on vacuum CF flanges. It is covered with thermal insulation made of extruded polystyrene (initially Perlite, an non-inflammable, expanded natural mineral was chosen, but was subsequently replaced; see discussion in Section\,\ref{sect:Screen}).

The vessel has been constructed such that each of the three volumes can be evacuated to -1.0\,barg independently. The main and the bath volume are tested up to +1.2\,barg, before the installation underground at LSC. After installing the TPC inside, the main volume is pumped for several months to reduce outgassing from detector components. For cryogenic operation, the pressure in the main volume is maintained at +0.1-0.2\,barg, to prevent air from leaking into the system. The main volume is protected ultimately with a rupture disk that opens at around +0.6\,barg, while an electro valve is set to release the pressure at +0.35\,barg. 

Regulation of the (vapour) pressure in the main volume is achieved by cooling the LAr through the vessel wall, by the sub-cooled LAr in the surrounding bath. The LAr in the bath is cooled by means of the cooling system, which is described in Section\,\ref{sect:Cooling}. Nominal operation parameters during \ardm\ \rI\ are summarised in Table\,\ref{tab:thermo_par}. 

\begin{table*}[htb]
\begin{center}
\begin{tabularx}{\textwidth}{@{\extracolsep{\fill}}lcc}
\hline
Volume                & Pressure/Vacuum [mbara] & LAr temperature [K] \\
\hline 
Main detector volume  & 1015                    & 87.3 \\ 
Cooling bath          & 870                     & 85.9 \\
Vacuum Insulation 1   & $<$2$\cdot$10$^{-4}$    & - \\
Vacuum Insulation 2   & $<$5$\cdot$10$^{-4}$    & - \\
\hline
Hall\,A                & 885                     & - \\
\hline
\end{tabularx}
\caption{Operation parameters of \ardm\ \rI\ with the pressures in the two cryogenic systems and the corresponding LAr temperatures. The mean values of the atmospheric pressure in Hall\,A is shown for comparison.}
\label{tab:thermo_par}
\end{center}
\end{table*}

\subsubsection{Vacuum system}
\label{sect:Vacuum}

The vacuum insulation of the experiment is subdivided in two separate volumes as shown in Figure\,\ref{fig:CryoDesign}: one around the detector vessel (right, Vacuum Insulation 1) and the other around the cryogenic services (left, Vacuum Insulation 2). The instrumentation of both vacuum systems is identical: (1) a gate valve, (2) a turbo molecular pump (TMP) and (3) a backing oil pump. For nominal operation, we keep the two volumes pumped continuously. The fail-safe gate valve steered by compressed air closes in case of electrical power outage, and maintains the vacuum. In addition an electrical safety valve and an oil filter are installed between the TMP and the backing pump. 

Each volume is equipped with a pressure sensor for the range 0.05-2\,bara and a thermal conductivity (Pirani) vacuum gauge for values down to $\sim$10$^{-4}$\,mbara. Sequences for starting and stopping the pumping, including opening/closing of the gate valve, starting/stopping of the backing pump and TMP, are programmed as fully automatic processes of the ArDM process control system, based on a programmable logic control (PLC) (see Section~\ref{sect:SC}). 

\subsubsection{Cooling system}
\label{sect:Cooling}

When bath and the main vessel are filled with liquid argon, the heat input from the environment is found to be $\sim$500\,W. A redundant cryo-cooling system consists of three Gifford-MacMahon cryorefrigerators, CRYOMEC AL300\footnote{\url{http://www.cryomech.com/products/cryorefrigerators/gifford/al300/}}, each having a cooling power of 266\,W at 80\,K with 50\,Hz AC electrical supply. Therefore to maintain the thermodynamic conditions stable, it is sufficient to keep two cryocoolers operating, keeping one spare. 
The regulation of the thermodynamic conditions is achieved by regulating heaters placed onto each cold head, controlled by a proportional-integral-derivative controller (PID controller) integrated in the PLC. The full cryogenic cooling power is used to maximise condensation during LAr filling, where GAr at room temperature from standard 200-bar bottles is cooled down and liquefied into the main vessel. A filling rate of $\sim$70\,L/day of LAr was achieved during the filling. 

\subsubsection{Argon purification system}
\label{sect:Purification}

The argon purification system is designed to provide sufficient cleaning power of the LAr target to exceed 1\mis\ for free electron lifetime, necessary for drift lengths of the order of 1\,m at an electric field of $>$0.2\kv/\cm. This translates into an oxygen-equivalent impurity level of 0.1\,ppb\,\cite{BUCKLEY1989364}. The experiment is equipped with two independent circuits to remove impurities trapped in both the liquid and the gaseous phases. In the sealed system of \ardm, the main source of impurities is outgassing from internal detector components. 

\paragraph{LAr purification system}
An internal cryogenic bellow pump recirculates the LAr of the main detector volume through a pure Cu-powder cartridge embedded in the LAr bath, at a speed of $\sim$150\,L/h. This provides the main removal of electro-negative impurities, such as oxygen. The purification filter is a custom-made cartridge containing activated copper grains to bind oxygen in the chemical reaction: ${\rm 2Cu} + {\rm O}_2 \rightarrow 2{\rm CuO}$~\cite{Epprecht2012diss}. The double-bellow pump was also specially designed and constructed. The LAr flow rate can be adjusted by the pump frequency in the range 0.4-3.5\hz. At the nominal frequency of 1\hz\ and a total displacement of 48\cm$^3$/cycle the flow rate reaches about 170\,L/h, i.e. it takes about 8 hours for one volume exchange of 1.4\,m$^3$ of LAr. 

\paragraph{GAr purification system} 
The GAr is recirculated by means of an external pump and a room temperature getter cartridge at a speed of $\sim$4000\,L/h. The pump is of a double diaphragm KNF pump\footnote{\url{http://www.knf.com}} of the type N 0150.1.2 AN.12 E. The gas is taken from the top of the detector via a 4\,m long heatable line and pumped through a commercial room-temperature getter of the type MicroTorr MC4500-902FV from the company SAES\footnote{SAES Pure Gas Inc.,\,\url{http://www.saespuregas.com}}. The return of the gas to the LAr system is done via a condenser immersed in the LAr bath. The temperature of the line is monitored by the PLC.

\subsection{Light readout system}
\label{sect:LRO}

To detect VUV scintillation light at LAr temperatures we adopted a design of WLS coated reflectors combined with borosilicate windowed cryogenic PMTs. The PMT windows are coated with WLS as well, to detect directly impinging VUV light. The active cylindrical LAr target volume is contained entirely inside an optically closed surface. The photocathode coverage amounts to approximately 70\% of the top and bottom readout planes, or $\sim$14\% of the total inner surface. 

\subsubsection{The wavelength shifter and the coating method}
\label{sect:WLS}

Among a range of wavelength shifting chemicals, the organic WLS material Tetraphenyl-Butadiene (TPB) is considered for its fast response caused by the rapid process of radiative recombination of electron-hole pairs at the benzene rings in their chemical structure. Such a feature is required for recording undistorted waveform of the fast scintillation component that decays with a time constant of the order of a couple of nanoseconds, which is essential for the pulse shape discrimination of the electron recoil backgrounds from nuclear recoil signals. After R\&D work on a range of organic WLS's such as p-Terphenyl, POPOP, PPO and bis-MSB\,\cite{Boccone:2009kk,Boccone:2010ada}, we choose TPB, 1,1,4,4-tetraphenyl-1,3-butadiene, for the best light yield in conversion of argon scintillation light and detection of the shifted visible light by bialkali photocathode. TPB is particularly well suited for the detection of VUV light due to the large Stokes shift. The fluorescence decay time is about 1.68\ns\ and, since no phonon is involved, the recombination process does not slow down significantly at cryogenic temperatures. TPB coatings can be made durable with good adherence to the substrate and some resistance to mechanical abrasion. The coatings are generally not soluble in water but can be removed when necessary by using toluene, chloroform (CHCl$_{3}$) or other organic solvents. TPB coatings have been exposed to high vacuum conditions for very long periods of time and show no evidence of significant change in the detector sensitivities. 

Different techniques have been tested for deposition of TPB on different substrates, and best results are obtained by the vacuum evaporation deposition method. The substrate to be coated is mounted in a vacuum chamber (evaporator) at a fixed distance above one or more crucibles filled with a certain amount of TPB. After pumping below 10$^{-5}$\,mbar by means of a TMP, the crucibles are heated slowly over several hours to about 220$^{\circ}$C to evaporate the TPB. TPB diffuses isotropically in a 2$\pi$ solid angle and is deposited by forming a molecular layer on the substrate surface. The layer thickness can be controlled by the amount of TPB filled in the crucibles. This technique is used for most of the TPB coated surfaces, i.e.\,the main reflector and PMT windows. Some less important components, e.g. the top/bottom reflectors bridging the space in between the PMTs, are dip-coated for convenience. In this case the substrate is immersed in a TPB organic-solvent solution and gently taken out and dried. TPB forms a layer on the surface but also is prone to crystallisation. 

\subsubsection{The WLS coated reflectors}
\label{sect:Reflector}

The reflectors are all based on PTFE material and are deployed at 3 different sections of the detector. The main reflectors, which line the side of the drift cage, are made of 254\mum\ thick and 20$\times$108\cm$^2$ large sheets of the PTFE fabric Tetratex\trt\ (TTX), produced by the company Donaldson Membranes, USA. A total of 13 such foils are used to entirely line the side walls of the drift cage. The TTX fabric is favoured against standard PTFE sheets due to the better adhesion of TPB. To mechanically support the soft fabric, a multi-layer plastic reflector film, Vikuiti{\texttrademark} ESR foil (Enhanced Specular Reflector) from the company 3M, USA, is sewed to the back side of the TTX sheets.

\begin{figure*}[htb]
\begin{center}
\includegraphics[width=0.9\textwidth]{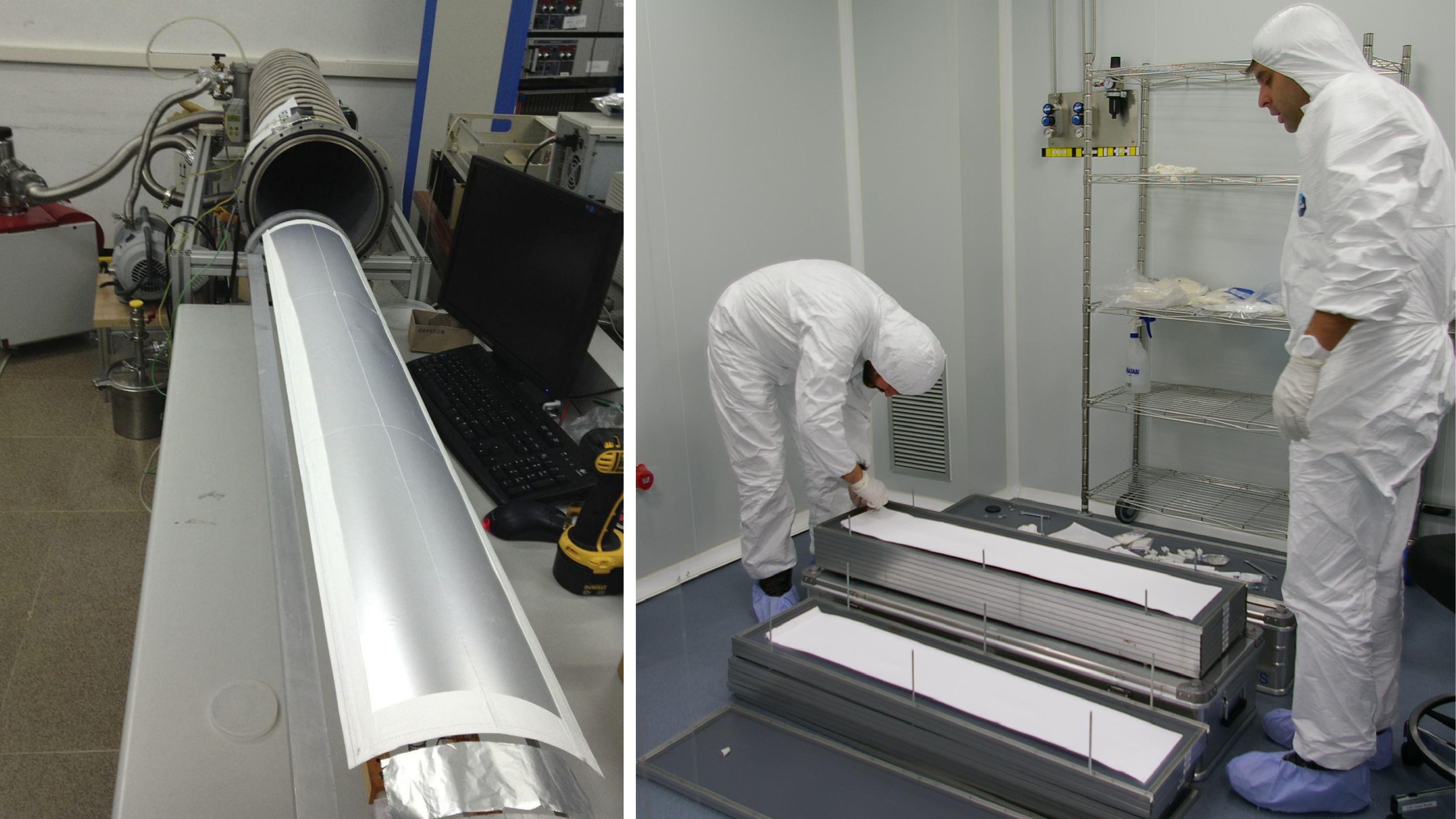}
\caption{Preparation of the main reflector foils. The evaporator is visible in the left figure.}
\label{fig:main_reflector}
\end{center}
\end{figure*}

The coating of the TTX sheets with TPB is done under vacuum evaporation deposition using a custom-made evaporator (Figure\,\ref{fig:main_reflector}). A layer thickness of 1\,mg/\cm$^2$ is chosen \cite{Boccone:2009kk,Boccone:2010ada}, which means total amount of TPB deposited on 20 reflector foils including spares is 43\,g while we evaporated total of 104\,g of TPB in the crucibles.

The top and bottom reflectors, covering the area between the PMTs, are made of 1\mm\ thick PTFE sheets. A large PTFE sheet is cut with water jet in a shape of a disc (900\mm\ in diameter) having 12 holes 20\cm\ in diameter where the spherical PMT windows are to be inserted. After the machining and cleaning the reflector sheet is dip-coated with TPB.

The third reflector type, bridging the conical transition ($\sim$15\cm) from the cylindrical section of the main reflector to the larger PMT mounting plate, consists of the same TTX foil as used for the main reflector, but is left uncoated in order not to convert VUV light produced outside the drift cage. With this third reflectors connecting between the two types as described above, the inner surface of the active LAr target volume is lined entirely with PTFE reflectors without gaps and overall more than 80\% of the reflector surfaces are coated with TPB.

\subsubsection{The PMT arrays}
\label{sect:PMT}

\begin{figure}[htb]
\begin{center}
\includegraphics[width=0.5\textwidth]{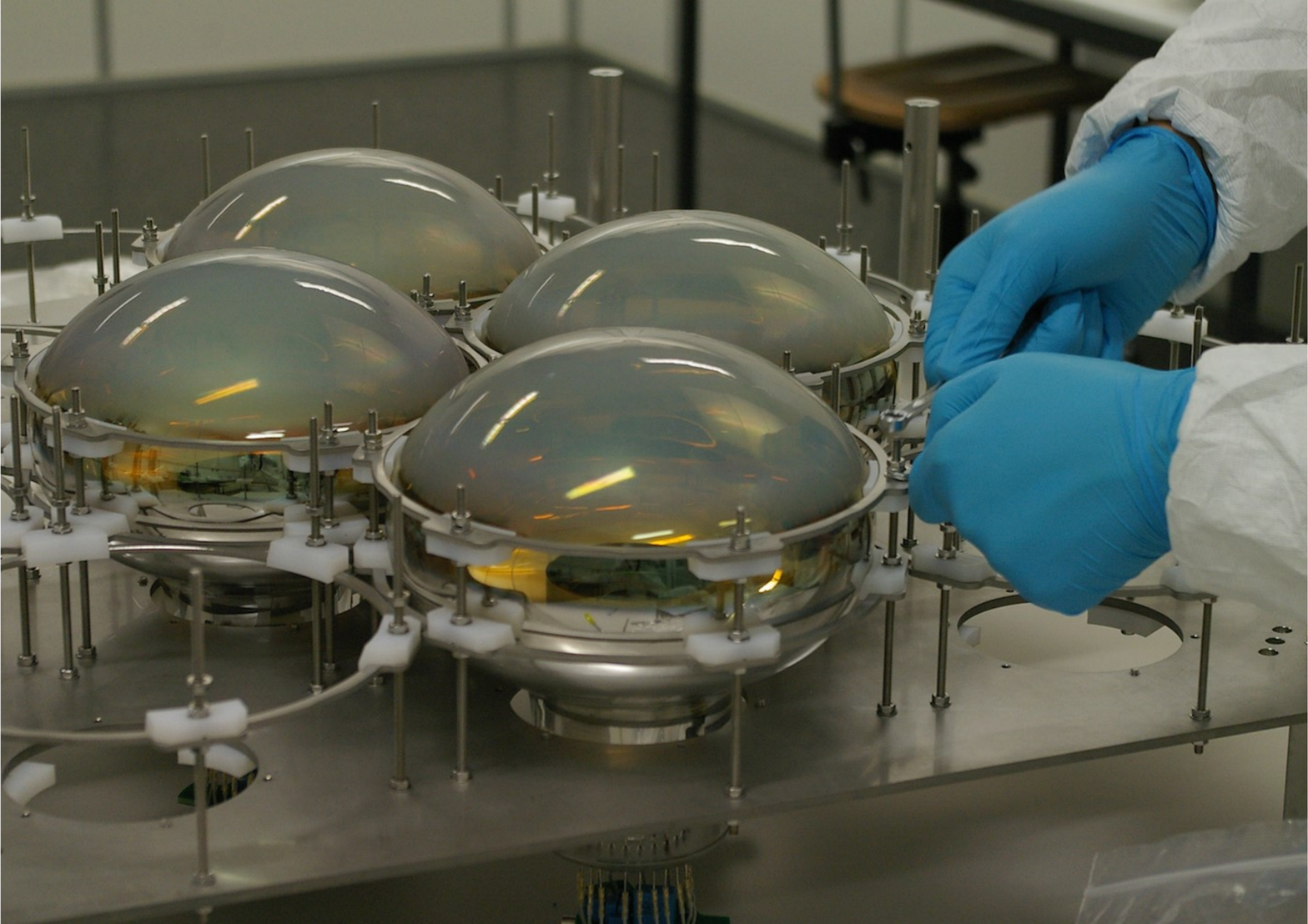}
\includegraphics[width=0.48\textwidth]{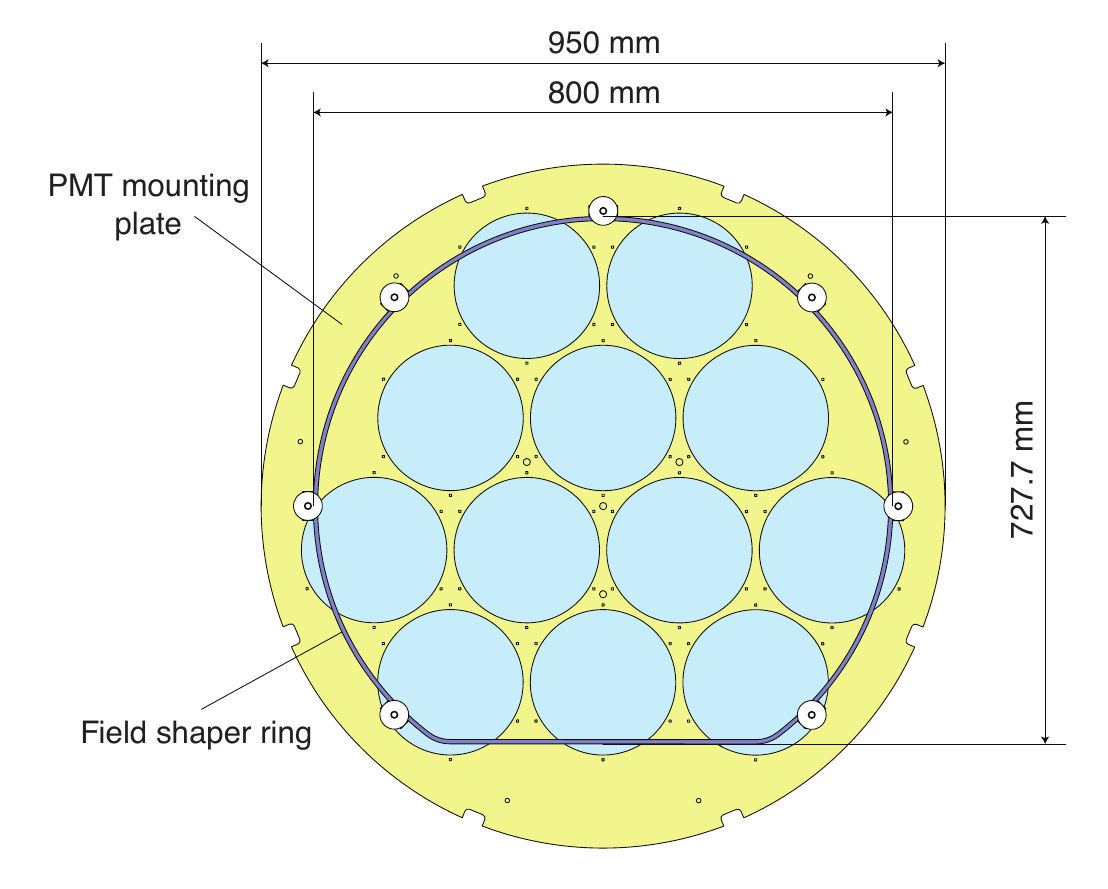}
\caption{Arrangement of the 12 PMTs in the readout planes on top and bottom.}
\label{fig:pmt_array}
\end{center}
\end{figure}

A total of 24 8'' cryogenic Hamamatsu R5912-02MOD-LRI PMTs made from particularly radiopure borosilicate glass are used to assemble two identical PMT arrays installed mirror symmetrically. Figure\,\ref{fig:pmt_array} shows the layout of the PMTs in an array. Such an arrangement in a triangular symmetry is chosen for the largest coverage and the best symmetry in light collection around the axis of the cylindrical target volume. 

On a common stainless steel base support plate, each individual PMT is held with an independent supporting structure consisting of 2 stainless steel rings and 12 small polyethylene (PE) pads.
The spherical part of the PMT glass is clamped by the PE pads. The supporting system is designed so that the thermal contraction in cryogenic temperatures of the stainless steel ring and the PE pads are compensated and that the displacement of the position of the contact to the PMT is virtually zero. 

The PMTs own bialkali photocathodes with a platinum (Pt) underlay to preserve electrical conductivity of the photocathode at cryogenic temperatures. An approximately flat quantum efficiency (QE) of $\sim$17\% is obtained in the range of 360-430\nm . Due to the Pt underlay a reduction in sensitivity of roughly 25\% has to be taken into account in comparison to typical bialkali photocathodes operated at room temperature. Each PMT features 14 dynode stages and the gain reaches 10$^9$ at an operation high voltage (HV) of 1.7\kv, as quoted by the manufacturer. The high gain feature helps in operating the PMTs at relatively low bias voltages, which is convenient in particular for the PMTs located in the pure argon gas where electric discharges can often be an issue. To match the front-end readout electronics a nominal gain of 5$\times$10$^7$ is chosen, resulting in the bias voltages in the range 946-1345\,V for 24 PMTs in cold operation. The voltage divider is made on an FR-4\footnote{FR-4 is a glass-reinforced epoxy laminate} printed circuit boards (PCB) using cylindrical metal film SMD resistors. A total resistance of 13.4\,M$\Omega$ is chosen to reduce the divider current and consequently the power dissipated in LAr to minimize bubbles creation. A typical operation HV of 1.1\kv\ leads to the divider current of 82\,$\mu$A and the dissipation of 90\,mW/PMT. In order to ensure a good pulse linearity, large capacitors are connected in parallel to the resistor chain at the last five dynode stages, i.e. 22\,nF for the dynodes 10 and 11, 47\,nF for the dynodes 12-14. Polypropylene capacitors are used for their reliability at LAr temperature. The divider circuit is designed paying particular attention to electric fields created on the PCB, in order to avoid electrical discharges.  

While a relatively thick layer of TPB can be used for the PTFE reflectors, coating on the PMT window requires a more precise optimisation. The coating should not compromise the detection efficiency for the visible light as a large part of the photons falling onto the PMT is already converted on the main reflector. The thickness of the TPB coating layer was optimised for the sensitivity to both, the VUV and the shifted blue light by tests in the lab,
scanning the performance of the TPB layer thicknesses in the range of 0.05-0.2\mgcs\ by measuring scintillation light of an \alp\ source in a gaseous argon test setup at room temperature. An increase of the efficiency is observed between 0.05 and 0.1\mgcs, while no significant difference is seen between 0.1 and 0.2\mgcs. Since no reduction of the efficiency to visible light was observed, finally we chose a layer thickness of 0.2\mgcs\ for the coating of the PMT windows. 

The TPB coating of the PMTs was performed at the Thin Film \& Glass Group of CERN, using its evaporator capable of coating one 8'' PMT at a time. In this setup the coating thickness is controlled by measuring mechanical oscillations of a crystal that is positioned beside the PMT, as the oscillation frequency changes as a function of the thickness of the deposited TPB layer. While the evaporation process takes less than one hour, the whole coating process requires a day per PMT mainly due to the evacuation of the evaporator taking several hours. 

\begin{figure}[htb]
\begin{center}
\includegraphics[width=0.9\textwidth]{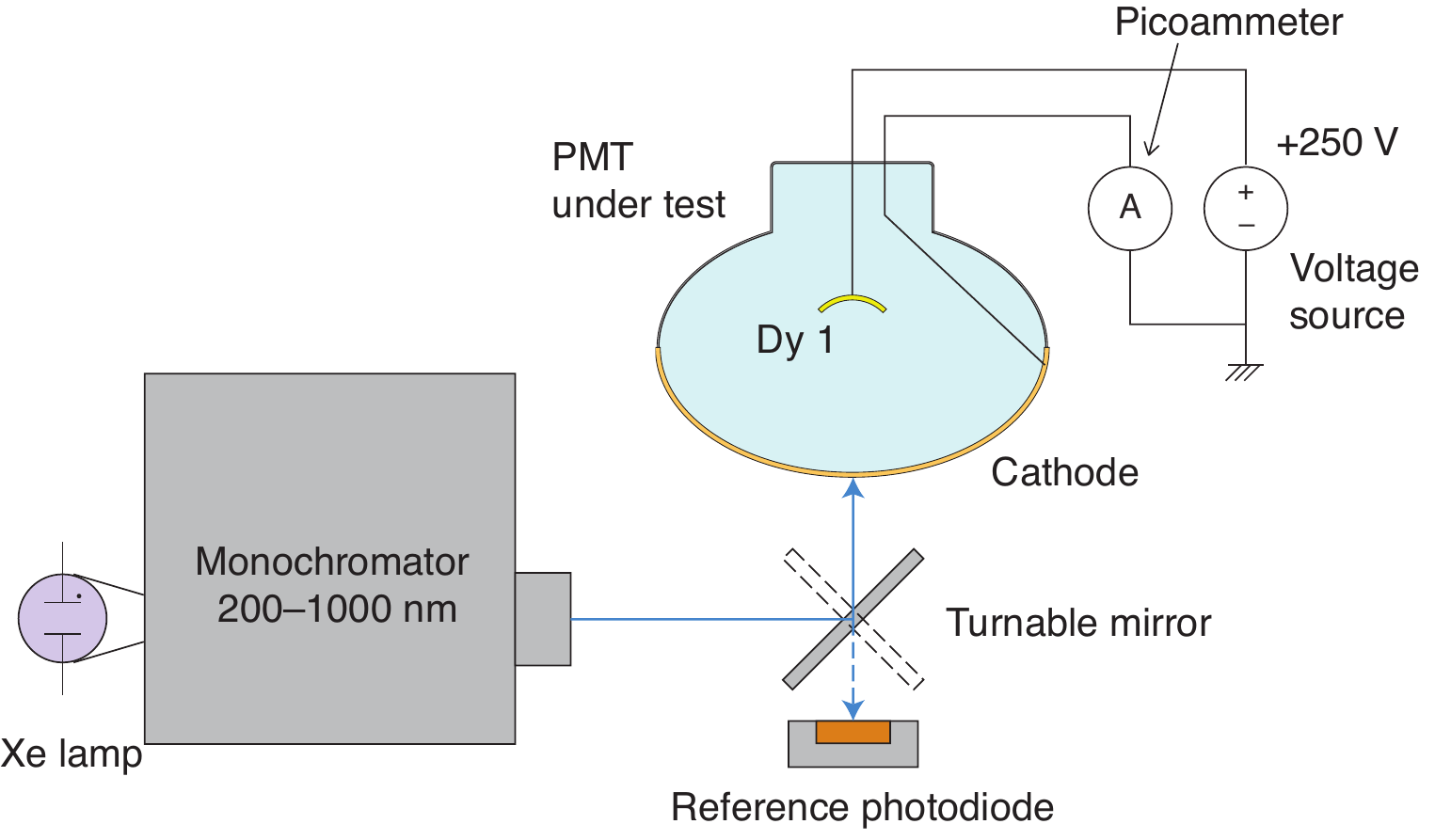}
\caption{Measurement setup to control the quality of the TPB coating applied to the PMT windows.}
\label{fig:qe_meas}
\end{center}
\end{figure}

The quality of the coating process is controlled in the laboratory at room temperature by means of a spectro-photometrical setup, consisting of a xenon lamp with diaphragm, a monochromator, a calibrated photodiode and a pico ammeter as illustrated in Figure\,\ref{fig:qe_meas}. The incident wavelength can be selected between 200 to 1000\nm\ at an adjustable step size. The photocurrent of the PMT under test is measured using the pico ammeter, applying a positive bias voltage of 250\,V to the first dynode to collect photoelectrons from the photocathode. Turning a mirror by 90$^{\circ}$ the same measurement can be  done with a reference photodiode with known QE. The QE of the PMT at wavelength $\lambda$ can be calculated using 
\[
QE_{\rm PMT} (\lambda) = QE_{\rm PD}(\lambda)\cdot\frac{I_{\rm PMT}(\lambda)}{I_{\rm PD}(\lambda)}, 
\] 
where $QE_{\rm PD}$ is the known QE of the reference photodiode, $I_{\rm PMT}$ and $I_{\rm PD}$ the measured currents of the PMT and the photodiode, respectively.

\begin{figure}[htb]
\begin{center}
\includegraphics[width=0.9\textwidth]{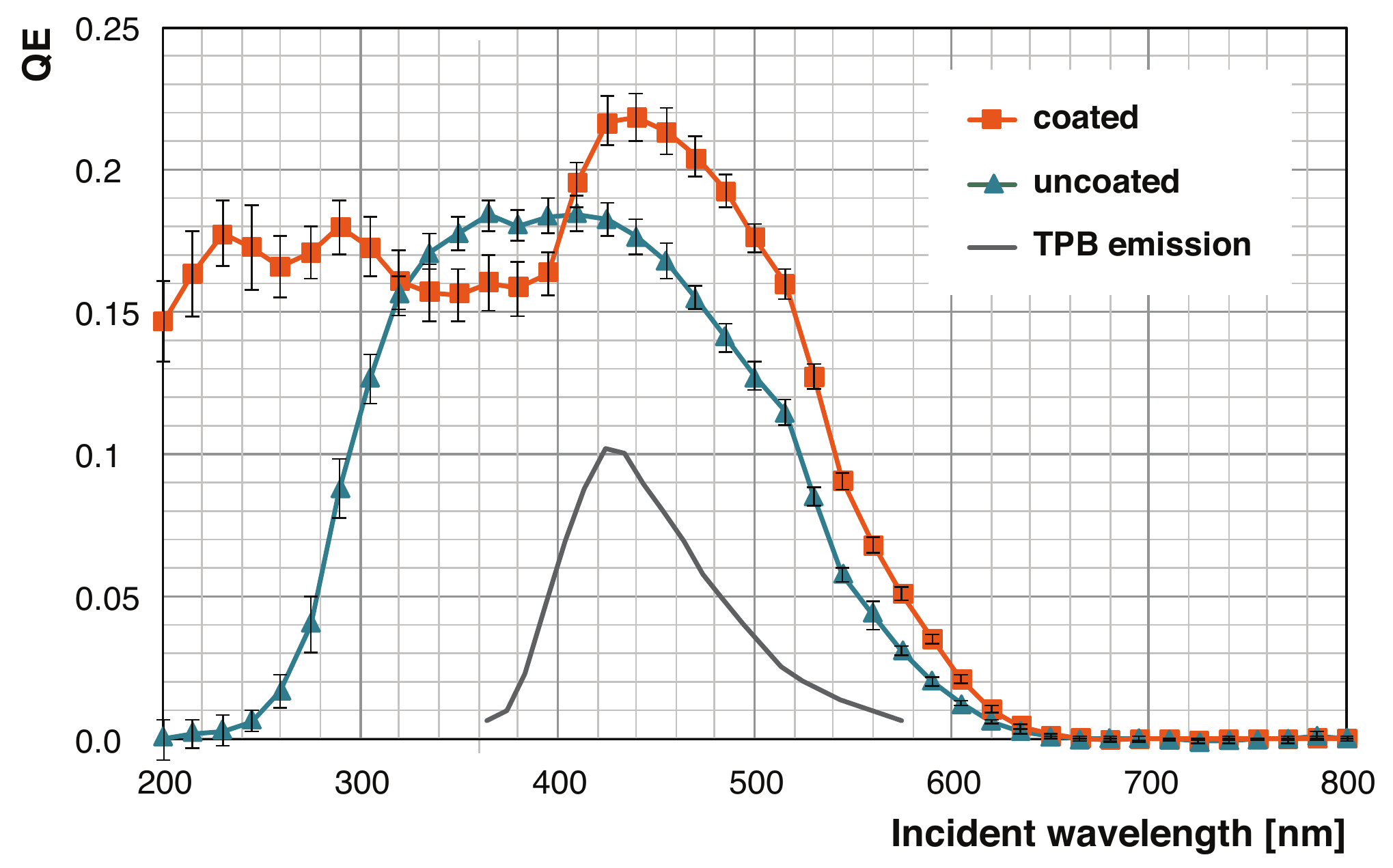}
\caption{QE spectra, averaged over 12 PMTs, before and after coating (see text).}
\label{fig:unco}
\end{center}
\end{figure}

Figure\,\ref{fig:unco} shows QE measurements averaged over the 12 PMTs of the bottom array. Triangles depict the QE curve before coating, while squares represent the curve after coating. In the UV region below 400\nm\ an approximately flat QE ($\sim$16.5\%) is obtained for the coated PMT demonstrating the effect of
the WLS emitting in 4$\pi$. Above 415\nm , where no shifting occurs, the about 15\% increase of the QE in respect to the uncoated PMT is due to scattering in the TPB layer and back reflection of scattered light onto the photocathode. For this reason the values reported here for QE are somehow biased by optical parameters of the measurement setup, but serve as a good handle to compare the performance of different coatings. However, from a simple solid angle estimate of the emitted shifted light we conclude a conversion efficiency of the TPB layer close to unity ($\sim$96.2\%).

\subsection{Background and shielding system}
\label{sect:BSS}

\subsubsection{Material screening} 
\label{sect:Screen}

Samples of detector components were screened for radioactive traces in the HPGe radiopurity screening facility of LSC (see Section\,\ref{sect:LSC}) at the underground site. The recorded \gam -spectra are analysed with emphasis on emission from the decay chains of $^{238}$U, $^{232}$Th, $^{235}$U, as well as from the individual isotopes $^{40}$K and $^{60}$Co. A list of screened components and their description is given in Table\,\ref{tab:screening-sum}. 
\begin{table}[htb]
\small
\begin{center}
\begin{tabular}{l l l c} 
\hline
Sample 	     &	Description 			                  	&	 Mass [kg]	  &	Time [d] \\
\hline	
PMT glass 	 & 	PMT - low radioactive glass (LRI)     &	0.7467	      &	45	\\
PMT metal	   &	PMT - Internal electrodes (metal)     &	0.197	        &	49	\\
PMT base 	   &	FR-4 boards - PMT bases 		          &	0.0366	      &	50	\\
SS struct	   &	Stainless steel - PMT support 	      &	2.077	        &	62	\\
SS clamp	   &	Stainless steel - PMT clamps	        &	0.2216	      &	36 	\\
SS rod	     &	Stainless steel - threaded rods    	  &	0.0606	      &	62	\\
PE clamp 	   &	HDPE - PMT clamps			                &	0.0632	      &	57	\\
PE shield	   &	External HDPE - Neutron shield	      &	0.8378	      &	33	\\
HVres 	     &	Ceramic HV resistors			            &	0.2427	      &	21	\\
Perlite      &	Perlite isolation material 		        &	0.1163	      &	45	\\
\hline 
\end{tabular}
\caption{Details on the screened material samples from inner parts of the \ardm\ detector.}
\label{tab:screening-sum}
\end{center}
\end{table} 

The screening campaign includes the PCB of the PMT voltage dividers, as well as various samples of HDPE. The specific activities (Bq/kg) of the materials are derived from the background subtracted \gam\ spectra taken at various high-purity Germanium counters of the screening facility. 

Among the items of main interest are the 8'' Hamamatsu R5912 PMTs as well as the innermost mechanical components of the detector. A GEANT4 Monte Carlo simulation is used to unfold effects from detection efficiencies and sample geometries of the screening setups. The method applied is described in\,\cite{Alvarez:2012as,Alvarez:2014kvs} and references therein. The decay chains of the isotopes of $^{238}$U, $^{232}$Th, and $^{235}$U studied in this work, are found, in fair agreement, to be in in equilibrium over their entire lengths. Their activities are determined from the weighted mean of identified \gam\ lines. In the case that no lines is found, the most stringent upper limit is used. For better comparison of the results radioactive activities are translated into contaminations in (kg/kg) and are listed in Table\,\ref{tab:screening-ppb}, where 1\,kru represents 10$^{3}$ decays/day/kg.

These results confirm our extrapolations of the material purity derived from literature or data sheets which are used for the construction of the detector. Table\,\ref{tab:screening-ppb} also lists the screening result for Perlite, originally planned to to be used for thermal insulating of the top part of the experiment. However due to its large contamination it has been replaced with extruded polystyrene. 
\begin{table}[htb]
\small
\begin{center}
\begin{tabular}{l l l l l c} 
\hline
Sample    	&	$^{238}$U [ppb]  &	$^{235}$U [ppb]  &	$^{232}$Th [ppb]  &	$^{40}$K [ppb]  & $^{60}$Co [kru] \\
\hline	
PMT glass 	&	51.7$\pm$0.3	   &	0.70$\pm$0.02    &	28.3$\pm$0.5	    &	1.7$\pm$0.07	  &	$<$0.2          \\ 	
PMT metal  	&	14.7$\pm$0.3	   &	0.71$\pm$0.04	   &	18.4$\pm$0.7	    &	12$\pm$0.4	    &	-	              \\
PMT base   	&	746$\pm$1	       &	9.0$\pm$0.1	     &	2720$\pm$10	      &	64$\pm$0.7	    &	-	              \\
SS struct  	&	0.257$\pm$0.002  &	$<$0.05		       &	1.57$\pm$0.01	    &	$<$0.04		      &	1.24$\pm$0.01   \\ 
SS clamp	  &	$<$0.6		       &	1.0$\pm$0.3	     &	$<$3 		          &	$<$0.1		      &	2.0$\pm$0.2     \\
SS rod	    &	$<$2		         &	1.18$\pm$0.08	   &	$<$6		          &	0.18$\pm$0.01   &	0.76$\pm$0.02   \\ 
PE clamp   	&	2.85$\pm$0.05	   &	$<$0.2		       &	23.3$\pm$0.6	    &	0.3$\pm$0.07	  &	$<$0.5          \\ 
PE shield	  &	0.34$\pm$0.06	   &	$<$0.03		       &	2.41$\pm$0.03	    &	0.06$\pm$0.01	  &	$<$0.06         \\
HVres 	    &	118$\pm$1	       &	1.92$\pm$0.02	   &	466$\pm$1	        &	6.7$\pm$0.06	  &	-	              \\
Perlite    	&	3650$\pm$20	     &	61$\pm$1	       &	13000$\pm$100     &	3400$\pm$47	    &	-	              \\
\hline 
\end{tabular}
\caption{Radioactive contaminations (kg/kg) of the elements used in ArDM listed in Table\,\ref{tab:screening-sum}, assuming the decay chains of the elements $^{238}$U, $^{235}$U and $^{232}$Th being in equilibrium.} 
\label{tab:screening-ppb}
\end{center}
\end{table} 

\subsubsection{Neutron shield}
\label{sect:Shield}

\begin{figure}[htb]
\begin{center}
\includegraphics[width=0.9\textwidth]{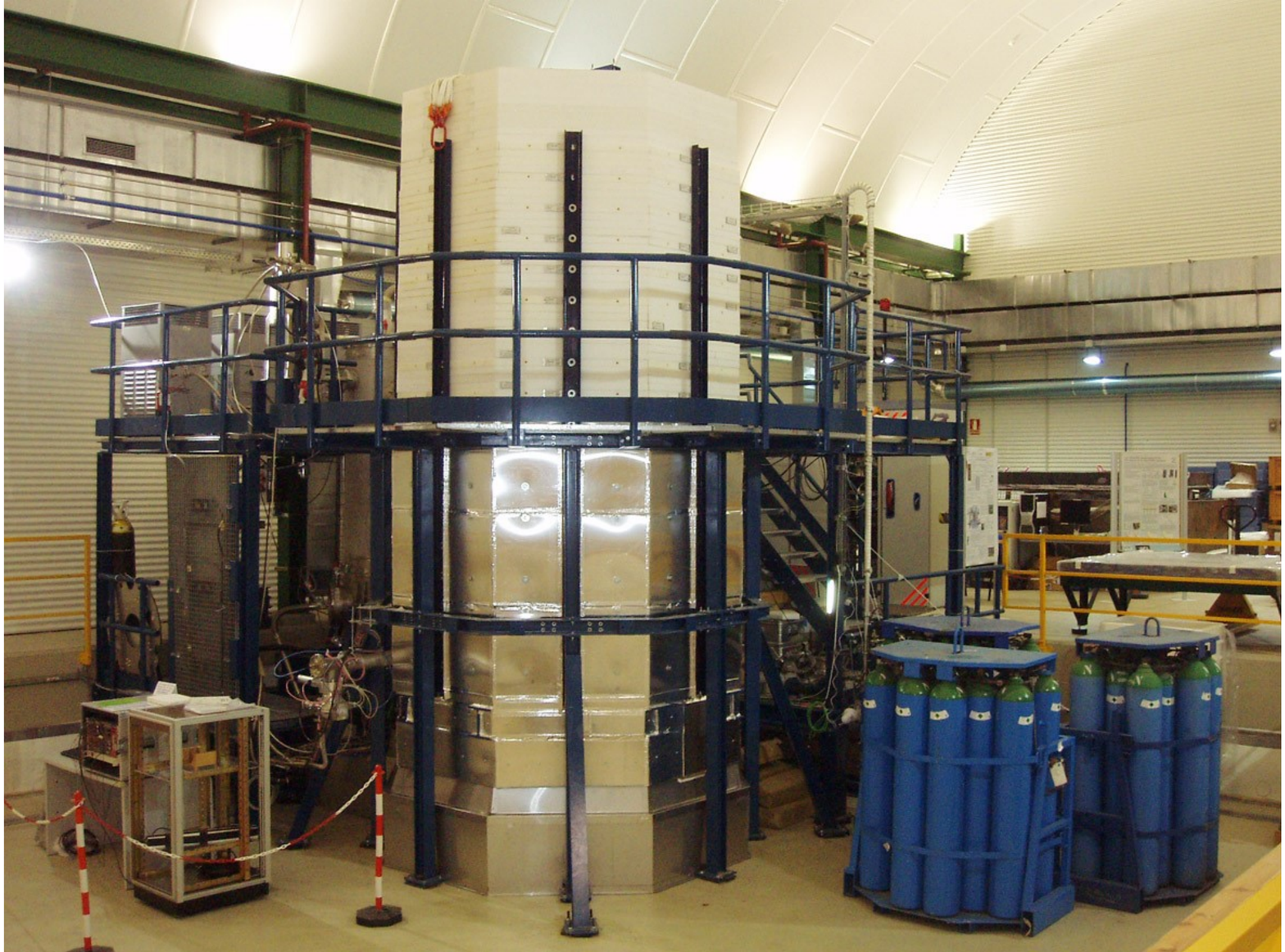}
\caption{Photograph of the mounted neutron shield made of HDPE.}
\label{fig:ArDMneutronshield}
\end{center}
\end{figure}

A passive neutron shield made of HDPE, fully surrounding the detector vessel, is built to reduce the neutron flux inside the detector volume coming from environmental sources such as surrounding rock of the underground lab. This shield structure has a shape of an equilateral octagon cylinder with a total height of 505\cm. See Figure\,\ref{fig:ArDMneutronshield}. The lateral part of the shield consists of a stack of octagonal ``rings'' of PE each one being composed of eight trapezoidal tiles.

The tangential outer and inner radius of the assembled ring are 121.2\cm\ and 71.2\cm\, respectively, resulting in a minimum wall thickness of 50\cm. The top cap is a pre-assembled unit having also a thickness (height) of 50\cm\ that can easily be opened and closed by crane. The bottom part is constructed underneath of the detector vessel integrated in the platform structure supporting the entire installation, having a maximum thickness of 72\cm\ to preserve 50\cm\ for parts having 12\cm\ high grooves to let horizontal beams of the platform pass through. To enclose the detector entirely with PE panels, two cylindrical 50-cm-long HDPE blocks are inserted in two horizontal pipes of Insulation Vacuum 2 connecting the detector vessel and the cryogenic service part. The total mass of the PE neutron shield amounts to $\sim$17\,tonnes.

The entire supporting structure, holding the 17\,tonnes PE neutron shield, is designed to be earthquake tolerant, free of resonances below 10\hz\ and sustaining horizontal accelerations on the order of 5\,m/s$^{2}$. To prevent the large mass of flammable PE from catching fire, the lateral surfaces of the shield are covered with fire protection panels, consisting of an aluminum sheet and a mineral wool thermal insulation layer. The top and bottom part is painted with fire retardant paint. 

Monte Carlo simulations based on flux and spectrum of neutrons in Hall\,A from surrounding rock suggest a reduction of the rate of neutrons hitting the inner detector volume of about 10$^{6}$. Of the neutrons passing through the shield, only 12\% have energies above 100\kev\ and hence are able to produce nuclear recoils above the threshold used for WIMP searches.

\subsection{Control and monitoring system}
\label{sect:SC}

The control and monitoring system serves to monitor as well as to manipulate the apparatus in a fail-safe and secure operational mode. The system is based on a PLC unit using the CERN designed process visualisation and control system (PVSS) and data acquisition (DAQ) framework for combining the data of the distributed monitoring systems and for their visualisation and storage~\cite{thesis_UrsinaDegunda}. Besides the sensors and hardware controllers, the installation also comprises redundant power sources, battery driven UPS systems as well as the main DAQ and data storage system.

More than 100 sensors are permanently read out and more than 50 actuators are controlled via analog and digital I/O modules. I/O lines can be adapted for different sensors and actuators. A high reliability of operation is achieved by the use of dedicated hard- and software. About 50 temperature sensors, 20 vacuum and pressure sensors, as well as 3 oxygen deficiency sensors deliver the main information of the status of the experiment. To control LAr levels in the bath and in the dewar, capacitive level meters are used. The processor module provide real time applications like the PID regulation of the cryocoolers. The monitoring and the control of the HV power supply for the PMTs are also integrated in the system. 

Any controls, e.g. pump or shutter valve operation, cryogenic cooling power or the PMT HVs are embedded in an interlock operational concept controlled by the PVSS framework. The software also allows for real time changes of stored parameters or settings to remotely adjust to the desired operational point. The system is able to send out alerts in case of changes of logic conditions or values passing defined limits. The \ardm\ control system is connected to the LSC alarm system, like the start of the emergency power generator in case of power cuts.

The status of the \ardm{} is continuously recorded by the PVSS software. The recorded data are backed up in a MySQL database on a separate Linux PC via Ethernet. The MySQL data is watched over internet by means of a dedicated web site. 

A crucial feature of the control system is the ability for remote control and monitoring. It provides the entry point for the remote operation of the experiment via a PC.   

\subsection{Quantitative risk assessment}
\label{sect:QRA}

Prior to cryogenic operation, a quantitative risk assessment (QRA) of the experimental setup has been performed by the  Scientific Research center DE\-MO\-KRI\-TOS\,\cite{Demokritos}, specialised in reviewing industrial installations like nuclear power plants. The QRA has been independely reviewed by an external company, \nier\,\cite{Nier}, Italy.
Both reviews have been done in close collaboration with LSC. 
The assessment included the simulation of accidents by solving three dimensional transient dispersion problems involving double-phase cryogenic leaks. Accidents were classified into three levels, where only the third one was found to represent considerable risk to personnel at the underground site. The frequency for such an event was estimated to be  4$\times10^{-5}$ per year\,\cite{ArDM_QRA}. A containment pool below the \ardm\ vessel was added, as well as an automatic gas extraction system. Completion of the safety instrumentation and procedures have been required as a final step before the start of full cryogenic operation of \ardm.

\section{DAQ, Trigger, and data reconstruction}

\subsection{DAQ system}
\label{sect:DAQ}

\begin{figure}[htb]
\begin{center}
\includegraphics[width=0.9\textwidth]{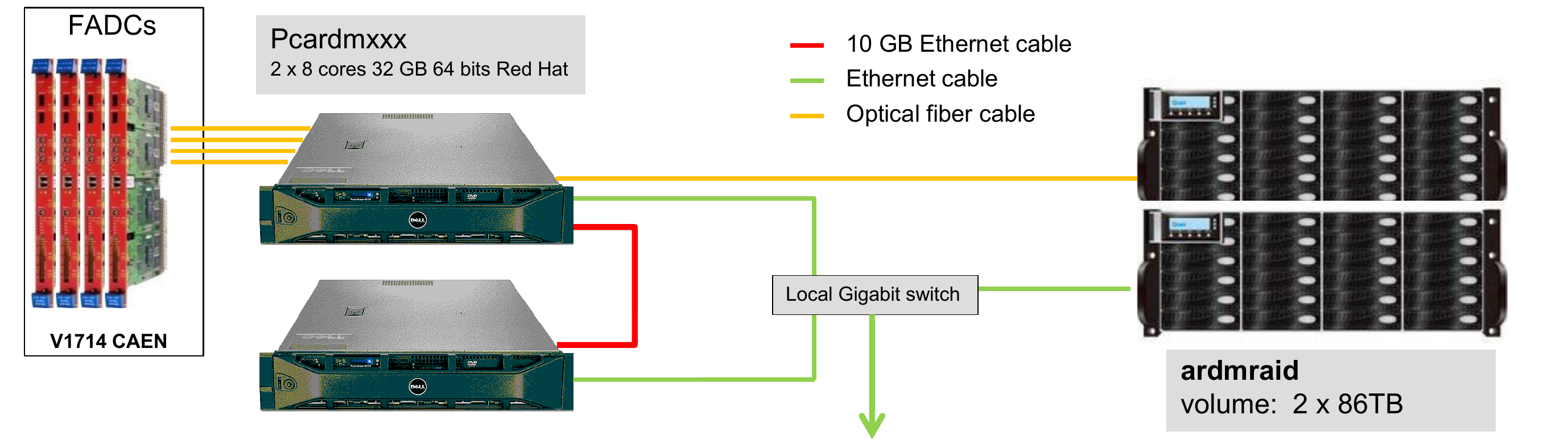}
\caption{Schematic diagram of the ArDM data acquisition (DAQ) system.}
\label{fig:daq_scheme}
\end{center}
\end{figure}

The DAQ system is designed and built to handle multi-kHz trigger rates and data rates up to $\sim$300\,MB/s to comply with the expected event rate from \ar\ events in an atmospheric argon target. Figure\,\ref{fig:daq_scheme} illustrates a schematic diagram of the DAQ. 
It consists of four 8-channel, 12-bit, 250\mhz\ FADC digitiser modules of the type CAEN V1720 for each of the PMTs, and was tested for trigger rates up to 3\khz. During \ardm\ \rI\ single phase configuration sampling time windows of 8 and 4\mus\ are used for gaseous and liquid data taking, respectively, resulting in 2048 and 1024 samples per PMT per event. Total event sizes are 96 and 48\,kB, respectively. The trigger rate is in the range of 1.3-2.1\khz\ depending on the stage of completion of the external PE shield. This corresponds to data rates of 85-138\,MB/s which could be handled dead time free by the DAQ system.

Data communications of the DAQ system uses optical links. Each of the four digitiser board hosting six PMTs is connected to the DAQ PC via a dedicated optical link, each having a maximum bandwidth of 80\,MB/s. The system thus theoretically is capable of handling total of 320\,MB/s with four parallel optical links. The DAQ PC has two CPUs each consisting of eight cores, and is equipped with two CAEN PCI-Express card each hosting four optical link ports. The DAQ software consists of two custom programs, i.e. {\it producer} for reading data from the digitiser and writing them in a shared memory, and {\it collector} for writing the data to hard disks. An independent set of {\it producer} and {\it collector} runs per optical link. These total of eight programs are controlled by another program, {\it manager}, and are distributed over eight cores of the CPUs. Data files are written individually for each optical link in the unit of sub-event that contains six PMTs, truncated into chunks of 1\,GB each. In order to reconstruct offline a full event from four independently recorded sub-events, internal clocks (TTT - Trigger Time Tag) of the four digitisers are synchronised and are reset at the start of the DAQ programs. An absolute time stamp of the event is recorded in each sub-event header with a resolution of the TTT-unit, i.e. 8\ns. 

The data storage system at the experimental site consists of 48 4-TB hard disks. Managed by a RAID6 controller, nearly 180\,TB is available for data storage. The storage system is connected via optical links (1\,GB/s bandwidth) to a dedicated DAQ PC, allowing for a writing speed of 500\,MB/s. Employing a lossless compression method (pigz - parallel gzip) the storage size is reduced by a factor of $\simeq 4$ and allows for continuous data recording of three months at a rate of 2\khz. Eventually, data is transferred via internet to the tape-based storage system CERN CASTOR\,\cite{CASTOR}. 

System monitoring is done by a dedicated software package for realtime data processing and online monitoring. Raw data is processed in parallel during data taking in a second DAQ PC connected to the system via an additional 1\,GB link. Its processing power can also be used to reduce storage size by close to real-time data sparsification or event filtering. For monitoring purpose the raw data is processed by the same software framework which is used for offline reconstruction, described in more detail in Section\,\ref{sect:RecoCal}. From data summary files the most relevant parameters, e.g. hit maps, noise rates, hit time distributions, synchronisation controls, energy spectra, spatial and temporal distributions and much more, are calculated and automatically broadcasted as graphs or histograms via a dedicated web server integrated into the control system. Status messages from the DAQ, as well as messages from experimental services are also displayed at this site. The control system includes an automatic watchdog unit with defined alarm levels sending out email and SMS messages in case of problems. 

\subsection{Trigger system}
\label{sect:Trig}
\begin{figure}[htb]
\begin{center}
\includegraphics[width=0.9\textwidth]{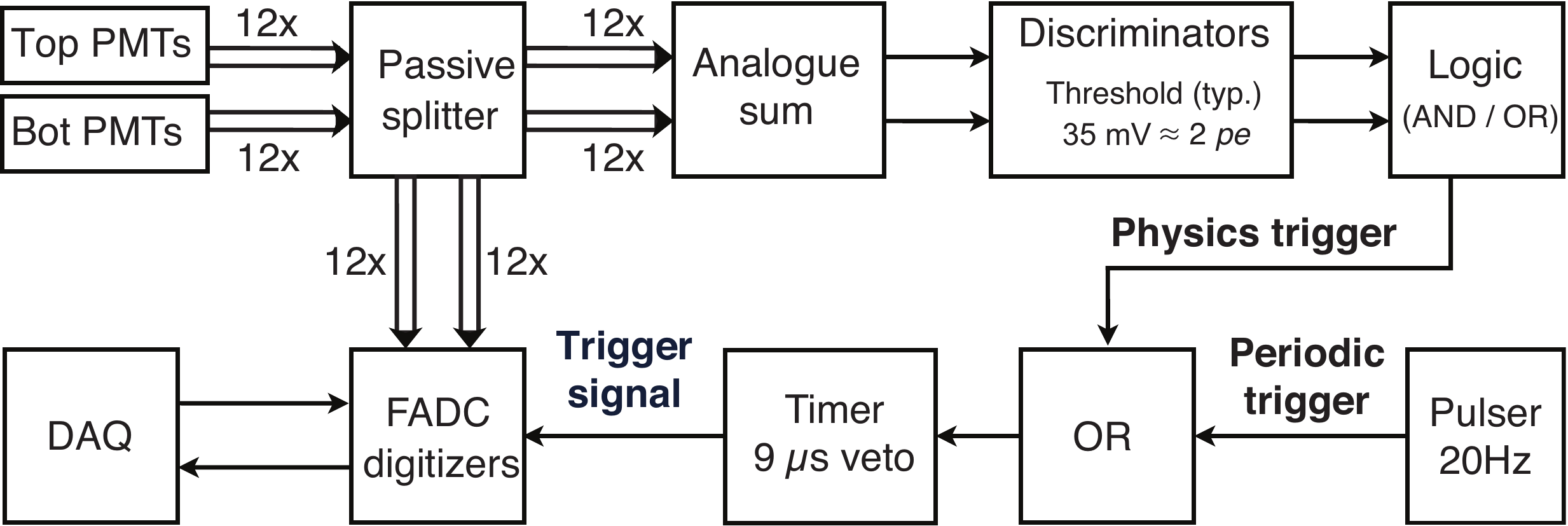}
\caption{Block diagram of the ArDM trigger system.}
\label{fig:trigger_scheme}
\end{center}
\end{figure}

The trigger signal is generated from both, the top and bottom PMT arrays. Figure\,\ref{fig:trigger_scheme} illustrates a block diagram of the trigger logic. Each of the 24 analogue PMT signals is split to equal parts into two, using a passive, resistive 50$\Omega$ divider. One output is fed to an input channel of a digitiser module, while the other is used for triggering. The entire trigger logic is made of CAEN electronics modules hosted in one VME crate. 

For data taking in \ardm\ \rI\ the {\it physics trigger} is created by either a logic AND or OR of signals from the top and the bottom array. A fixed discriminator threshold of 35\,mV is applied to the analogue sum of the 12 PMTs in each array, corresponding approximately to 2\,photoelectrons ($pe$) in the prompt light. After a trigger, new triggers are vetoed for a duration of 9\mus. This results in a dead time of less than 1\%.   

For general monitoring purpose (pedestals, noise, dark counts...) a periodic calibration trigger is generated using a pulse generator at a constant frequency. Such a {\it generator trigger} is added during the data taking. The generator frequency is set to 20\hz\ with the physics trigger rate of 1-2\khz\ and the fraction of the generator trigger events is around 1-2\% of all the triggers. The generator trigger can also be used for LED pulse calibrations of the PMTs. 

For \rI\ data taken at cold (cold gaseous or liquid target) the OR mode of the trigger was used thanks to a low dark count rate of the PMTs at low temperature (see Section\,\ref{sect:Cal}), while the AND mode served for first test runs with the gaseous target at room temperature.

\subsection{Event reconstruction}
\label{sect:RecoCal}

\begin{figure}[htb]
\centering
\includegraphics[width=0.9\columnwidth]{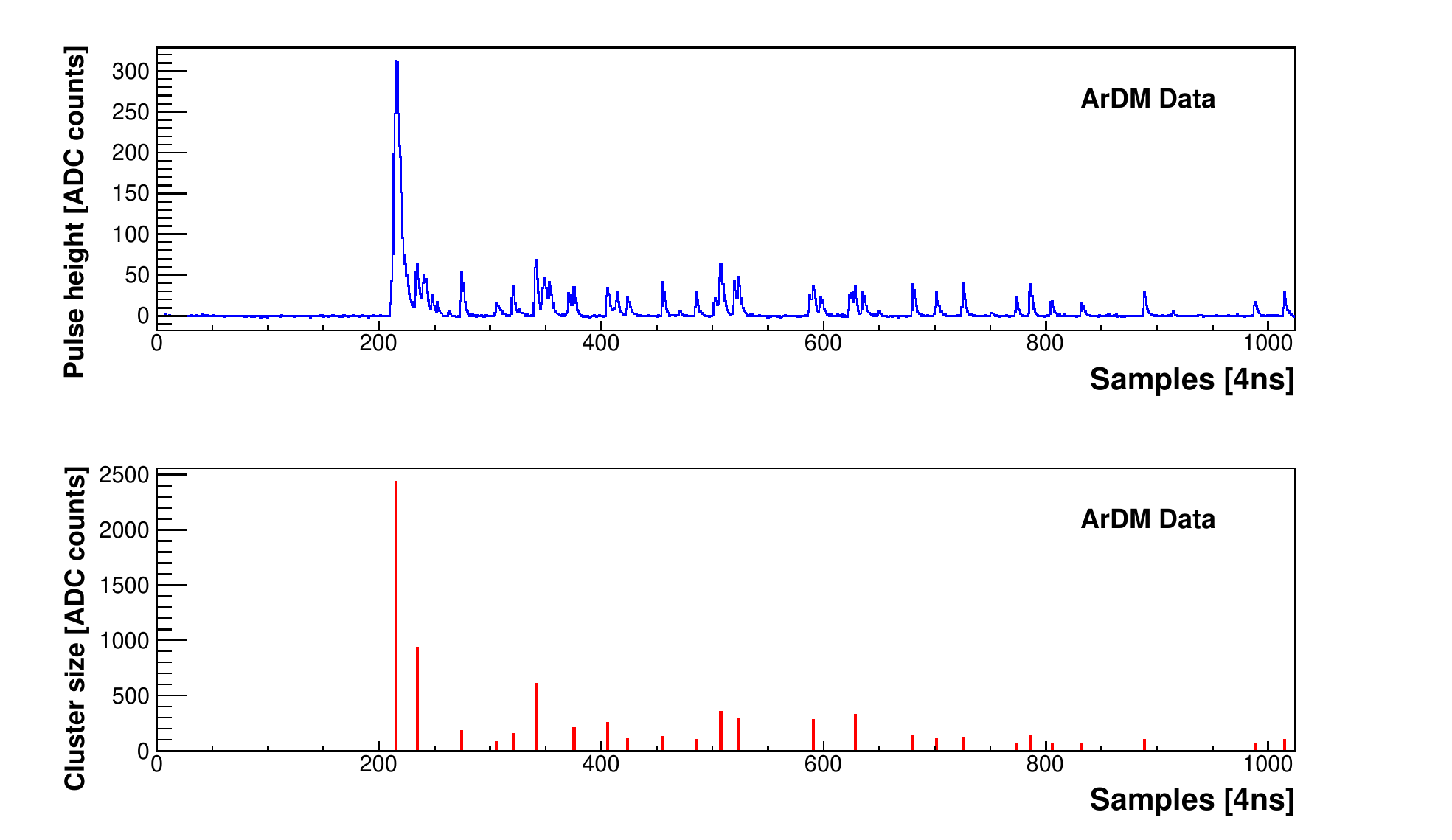}
\caption{Typical argon scintillation waveform, before (upper) and after (lower) hit-finding.}
\label{fig:pulse}
\end{figure}

The reconstruction of the scintillation light signal is done individually for the 24 waveforms of the PMTs by a dedicated software framework including several levels of processing. Main tasks include pedestal calculations, cluster or hit finding, and signal calibrations. The pedestals are evaluated from the first 200 samples (800\ns) in each PMT trace in the time region before the trigger by an iterative scanning algorithm to remove noise from dark counts or pile up. The mean single photo electron ($SPE$) charges are used for calibration. Their determination from signals in the event tails is described in Section\,\ref{sect:Cal}. 
Event reconstruction is driven by accurate photon counting. This is achieved by calculating the sizes ($S_{i}$) and times ($T_{i}$) of signal clusters or hits originated in the detection of individual or several (merged) photons. Cluster sizes are determined by summing adjacent samples above a threshold of 0.2\,$SPE$ in pulse height, corresponding to about 10\sig\ of pedestal fluctuations. Samples at the edges of signal clusters below this threshold are included, if their values are above 5\% of the maximal peak height of the cluster. In the case that two or more photo-electrons are close in time, the individual signals are merged into one cluster or hit. All quantities for further data treatment are calculated from the clusters or hits found in the 24 channels, as well as from a sum of all, the so called $virtual PMT$. Figure\,\ref{fig:pulse} shows an example for the waveform of argon scintillation event on one PMT before and after cluster (hit) finding. The reconstructed quantities are stored in ROOT tree structured files consecutively numbered in the same way as the raw data files, containing also the same number of events ($\sim$100k).

All analyses described here are based on summing the reconstructed signal clusters over corresponding PMT channels and time ranges. E.g.~the top and bottom light yields, $L_{\rm top}$ and $L_{\rm btm}$, are calculated from the sum of all signal clusters found in the top and bottom PMTs, respectively. The total detected light is calculated from the sum of all hits $L_{\rm tot} = L_{\rm top} + L_{\rm btm}$. A vertical localisation parameter $TTR$ is found from the ratio of the top to the total yield, $TTR = L_{\rm top} /L_{\rm tot}$. We use the $TTR$ value to estimate the vertical location of an event. 

\subsection{Calibration}
\label{sect:Cal}

Data is calibrated in two steps to reconstruct the deposited energy from an event which is related to the measured charge from the PMTs. Firstly the calibration of the single photon response is derived individually for all 24 PMTs from spectra of the integrated cluster charge of signal pulses from event tails or dedicated light pulse runs. Secondly, the light yield calibration is obtained by means of radioactive isotopes, i.e.\,the \ar, or by internal or external calibration sources, like \kr\ or \co. 

\paragraph{Single photon calibration} 
The individual PMTs have their HV adjusted to obtain a gain around 5$\times$10$^7$, so that each PMT contributed approximately equally to the trigger defined by the fixed discriminator threshold of 35\,mV (or approximately 2\,$pe$). The charge recorded by the FADC for a single photoelectron with this gain is 4\,pC after the one-to-one passive splitter. Terminated with 50\ohm, this equals to 0.2\,nV$\cdot$s or about 100 units for the integrated ADC counts. Taking typically three calibration runs where the HVs of all the PMTs are changed simultaneously, optimal HV values are determined from the fit to the gain curve obtained for each PMT. The single photoelectron responses over the 24 PMTs are adjusted within $\pm$4\% of each other.

For the purpose of monitoring and later analyses, a calibration constant for each individual PMTs is determined from the single photoelectron peak position in histograms of cluster sizes, on a run-by-run basis as part of the primary data reconstruction. Histograms of the integrated cluster charge are produced for 20  bins in time of the acquisition window for each PMT. The single photoelectron pulses are found from the time bins corresponding to the tail of the slow scintillation light. After rejecting the time bins where the mean of the distribution is larger than two photoelectrons, arithmetic means of the modes of the remaining histograms are computed and taken as the calibration constant. Such a method without using fits provided a robust, automatic evaluation of the calibration constants during the long-term data taking.

\begin{figure}[htb]
\begin{center}
\includegraphics[width=0.6\textwidth]{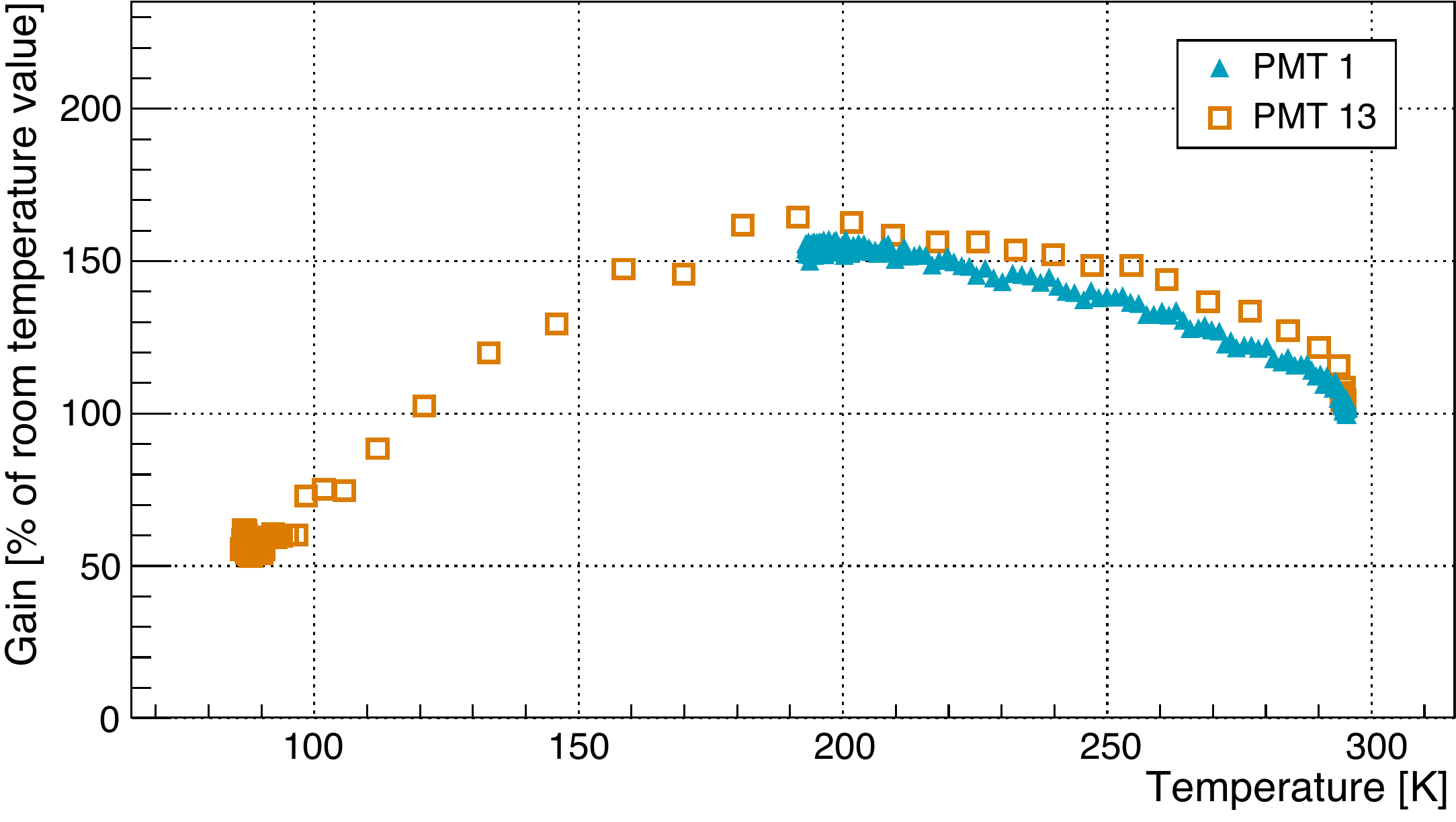}
\caption{Measured temperature dependence of the gain of the Hamamatsu R5912 PMTs at 1400\,V.}
\label{fig:PMTgainvsT}
\end{center}
\end{figure}

\paragraph{PMT gain} 
The temperature dependence of the PMT gains, as well as the PMT dark count rates, are measured during the gradual cooldown of the detector. The PMT gains, which depend on small changes in the work function of the dynode material, are shown for one of each PMT from the top (PMT\,1) and bottom (PMT\,13) array in Figure\,\ref{fig:PMTgainvsT}. Starting from room temperature, the gain, initially set to a value of 10$^{7}$ at 1400\,V bias, is found to increase until reaching a
maximum of about 160\% at 200\,K before it starts to decrease with further falling temperature. At the temperature of LAr (about 87\,K) the gain drops to about 60\% of its value at room temperature if the bias voltage is left unchanged. Due to its vicinity to warmer parts in the top of the experiment, PMT\,1 reached only about 170\,K in this first cool down. After the (later) filling of the target with LAr the temperature of the top array PMTs also approaches the value of LAr more closely. The gains of the PMTs are calibrated periodically to the standard value of 10$^{7}$ by changing the individual HV settings.    

\paragraph{Dark count rates} 
The rates of thermally emitted electrons from the photocathodes of the PMTs are determined from data taken with generator trigger superimposed to the physics trigger of LAr scintillation events. Here we describe a set of data (332k events) taken once the detector is under cryogenic conditions. The data is analysed for the number of signal clusters (mostly single photoelectron size) in each of the 24 PMT traces over the acquisition window of 4\mus. Random coincidences with LAr scintillation signals are removed by rejecting events with more than one signal cluster in a 48\ns\ window in the combination of top and bottom PMTs.

\begin{figure}[htb]
\begin{center}
\includegraphics[width=0.49\textwidth]{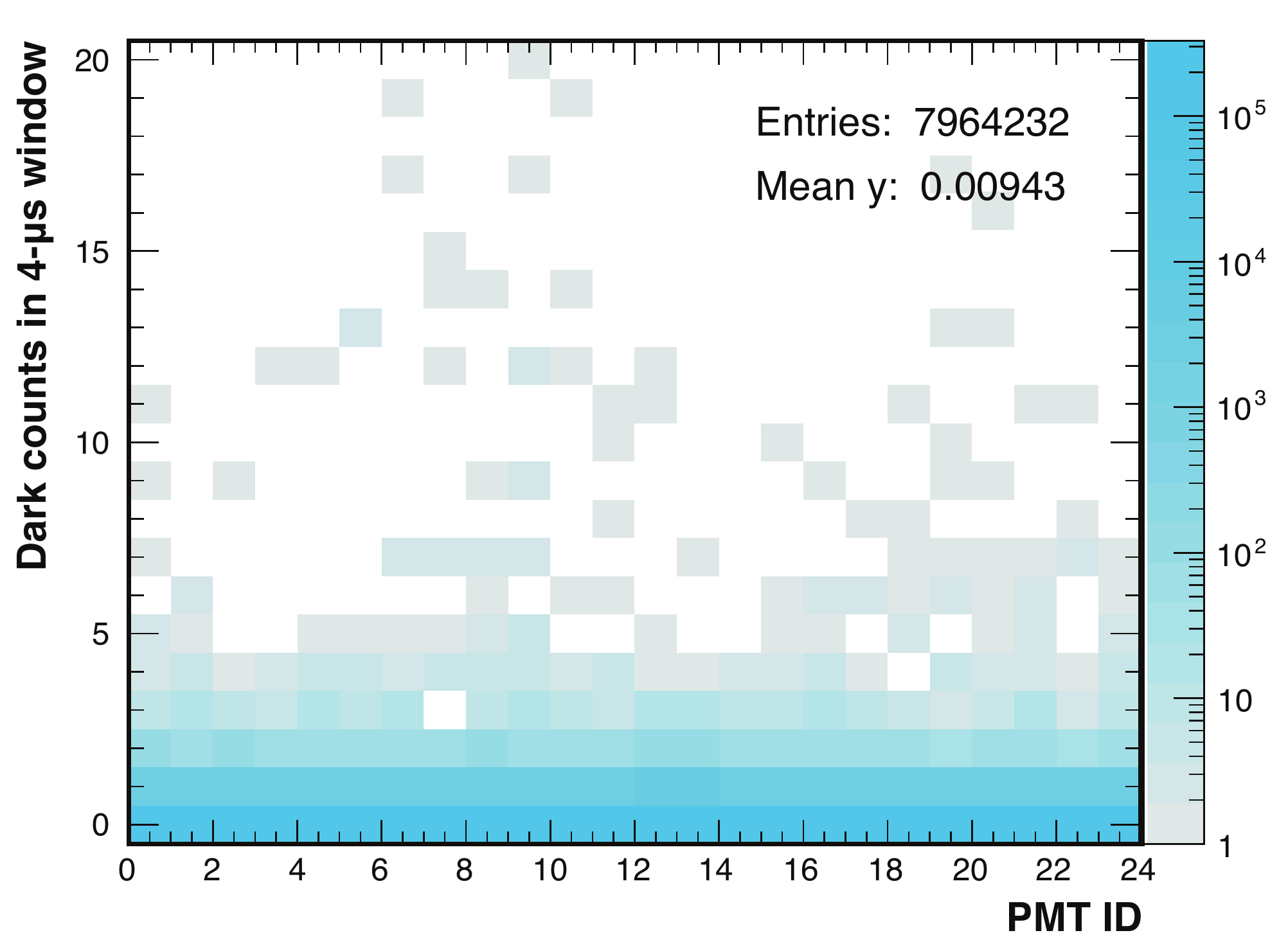}
\includegraphics[width=0.5\textwidth]{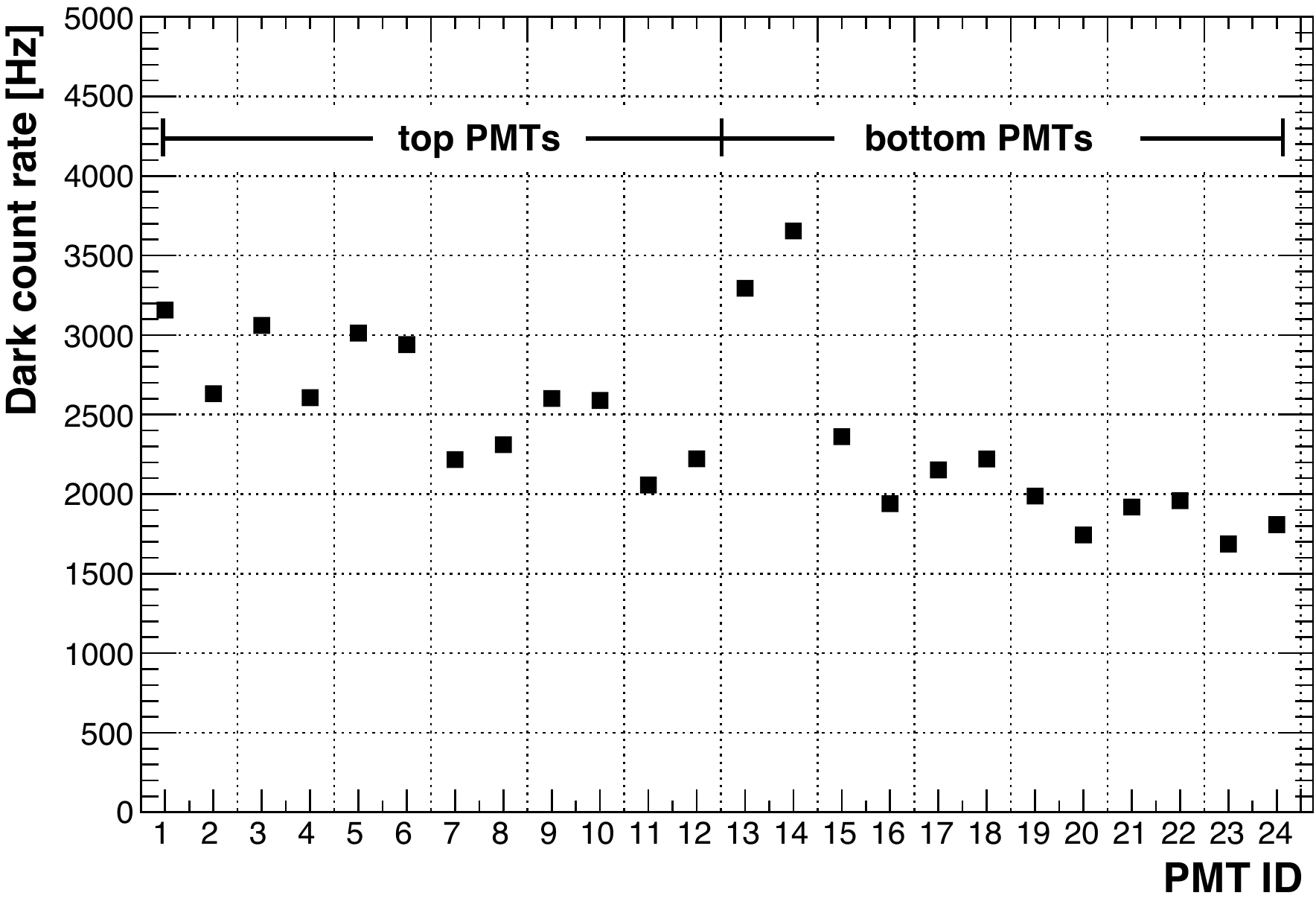}
\caption{Distribution of the number of dark counts within the 4\mus\ acquisition window for the 24 PMTs (see text). b) PMT signal rates determined from events triggered by the pulser.}
\label{fig:PMTdarkcounts}
\end{center}
\end{figure}

Figure\,\ref{fig:PMTdarkcounts} on the left shows the cluster count distributions for individual PMTs and on the right for all 24 PMTs, both in logarithmic scales. Final dark count rates are evaluated by Poissonian fits to the distributions to the left and are plotted to the right in Figure\,\ref{fig:PMTdarkcounts}.
The $\sim$20\% higher rate for the PMTs in the top array is due to a slightly higher temperature in the gas phase. It can also be observed that PMT\,13 shows a higher rate which is likel due to a self emission of small quantities of light, also cross talking to its nearest neighbor. Averaging over all 24 PMTs, we obtain a value of 2.3\khz\ per PMT at LAr temperature, which is found to be stable over time. The measured rate corresponds to only about 1\% probability per 8'' PMT to show one dark count during the acquisition time of 4\mus. A contribution from dark counts to the integrated pulse height of reconstructed scintillation signals can hence be neglected, which is an important property for the detection of LAr scintillation light with its long triplet decay time.  

\section{Detector commissioning and first results}
\label{sect:Results}

After the final assembly of the ArDM detector at the underground site, the main target volume was thoroughly evacuated at room temperature to remove outgassing material, above all H$_2$O molecules, which tend to stick to surfaces. Due to temperature-sensitive detector components the system could not be baked despite a general fully metal-sealed design. A vacuum of 10$^{-5}$\,mbar could be reached after several weeks of pumping. The experiment was then commissioned with argon gas of the type ALPHAGAZ\,2 (99.9999\% purity) from 200\,bar cylinders provided by Air\,Liquide, Spain.

\subsection{Detector response calibration sources}
\label{sect:Calsrc}

\paragraph{Internal low energy calibration source} 
A 100\,kBq $^{83}$Rb source was installed in a bypass circuit on the gaseous recirculation line, producing metastable \kr\ atoms. The de-excitation of \kr\ proceeds via two consecutive electromagnetic transitions, suppressed in their rates by large changes in nuclear spin. The first (32.1\kev) owns a half life of 1.8\,hours followed by a $\sim$150~ns delayed 9.4\kev\ transition. Most of the de-excitation energy, in total 41.5\kev, is used to eject electrons from the Kr atom via internal conversion and Auger effect. Due to the low range in the energy and short lifetime, \kr\ sources are well matched to the needs of highly sensitive  detectors in low background environments. In the ArDM experiment radioactive \kr\ atoms can be swept on demand into the gaseous phase of the main detector volume via the gaseous recirculation circuit. 

\paragraph{External calibration sources} 
Other radioactive sources were applied by placing them at different positions close to the outside of the main detector vessel. For this purpose some of the PE bricks of the neutron shield can be removed temporarily. In addition two plastic tubes were installed inside the neutron shield serving as guides for vertical and horizontal scans. Mainly a 180\,kBq \co\ (122\kev\ \gam) source was used for external calibration. Further-on the following sources were prepared for their use, but are not described in this work: 37\,kBq $^{22}$Na (0.511 and 1.27\mev\ \gam) a 37\,kBq $^{60}$Co (1.17 and 1.33\mev\ \gam) as well as a 20\,kBq $^{252}$Cf neutron source.

\subsection{Data with room temperature GAr target}
\label{sect:GasDataWarm}

\begin{figure}[htb]
\centering
\includegraphics[width=0.8\textwidth]{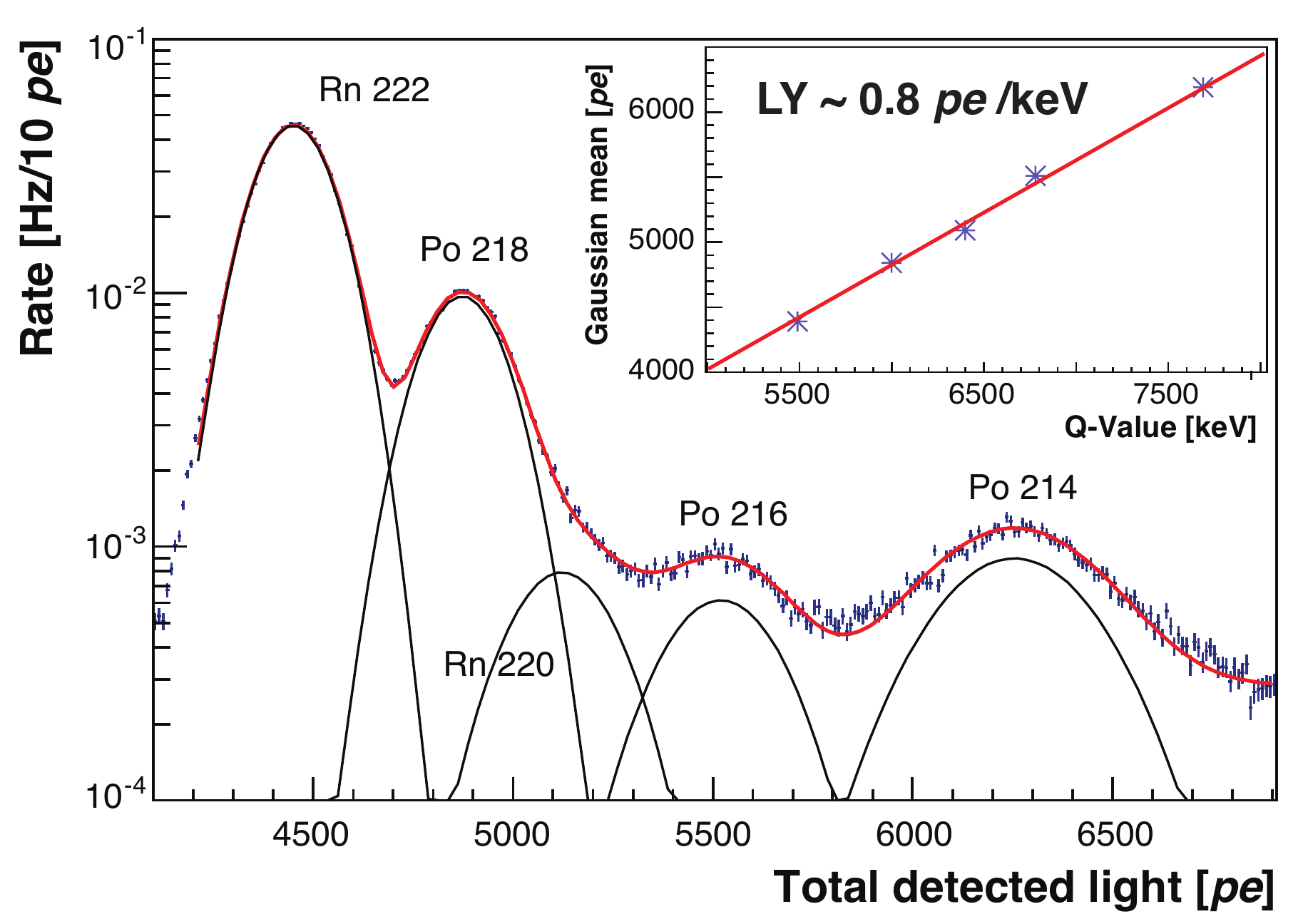}
\caption{Spectrum of total detected light in the large signal region for room temperature GAr data. The inset shows the light yields ($pe$) of the 5 fitted \alp\ lines.} 
\label{fig:AlphaGArData}
\end{figure}

The \ardm\ detector was commissioned firstly at room temperature with a GAr target over a period of about 3 month. The basic functionality of the detector components was verified and found to work as expected. In particular the cleaning efficiency of the recirculation system was confirmed by the increase of the slow scintillation time constant from the initial value of 2.5\mus\ after filling, to 3.2\mus. This value is close to the undisturbed lifetime of the triplet excimer state. However, a contamination of the argon gas with \alp\ emitters, was observed after activating the recirculation system, causing an increase of the trigger rate from about 20 to 30\hz. When the recirculation was stopped, a decrease of the \alp\ rate with a time constant of about 3.5\,d was observed, confirming the presence of $^{222}$Rn in the detector. While the presence of $^{222}$Rn should be avoided in a dark matter detector due to production of long lived radioactive isotopes, here it could be used for the light yield study of spatially uniformly distributed events in the high energy region. Figure\,\ref{fig:AlphaGArData} shows a fit (red) of the total detected light spectrum of GAr data with 5 (Gaussian smeared) \alp\ lines on a small background. We interpret these events as daughter products from Rn isotopes generated from $^{238}$U and $^{232}$Th decay chains in the SAES recirculation cartridge (see Section\,\ref{sect:Purification}). From these and other test source measurements (\co, \kr), we confirmed the good linearity of the light detection system (inset in the figure) and a yield of roughly 0.8\,$pe$/\kev\ for room temperature GAr. 

\subsection{Data with cold GAr target}
\label{sect:GasData}

Following the measurements in warm gas, the experiment was cooled down for operation with a cold GAr target. Of main interest was the functionality of the cryogenic system, as well as the performance of the light detection system at low temperatures.

The cool down of the setup was accomplished over a period of roughly one week by gradually filling the volume of the outer bath (see Section\,\ref{sect:CryoDesign}) with LAr. The argon gas in the main detector volume was kept at a pressure below the outer bath to prevent the formation of LAr in the main target. In this way temperature gradients and material stress on the detector components were also minimised. This method allowed for continuous monitoring of the detector functionality since the light readout could be kept running at all times. The temperature dependent gain curve of the PMTs (see Figure\,\ref{fig:PMTgainvsT}) was obtained from data taken during this period. Once the detector was in thermal equilibrium, the PMT gains were adjusted. The trigger rate was found to have increased to about 40\hz, explained by the 3.5\,times higher density of the argon gas at 87\,K in comparison to room temperature.

\begin{figure}[h]
\centering
\includegraphics[width=0.95\textwidth]{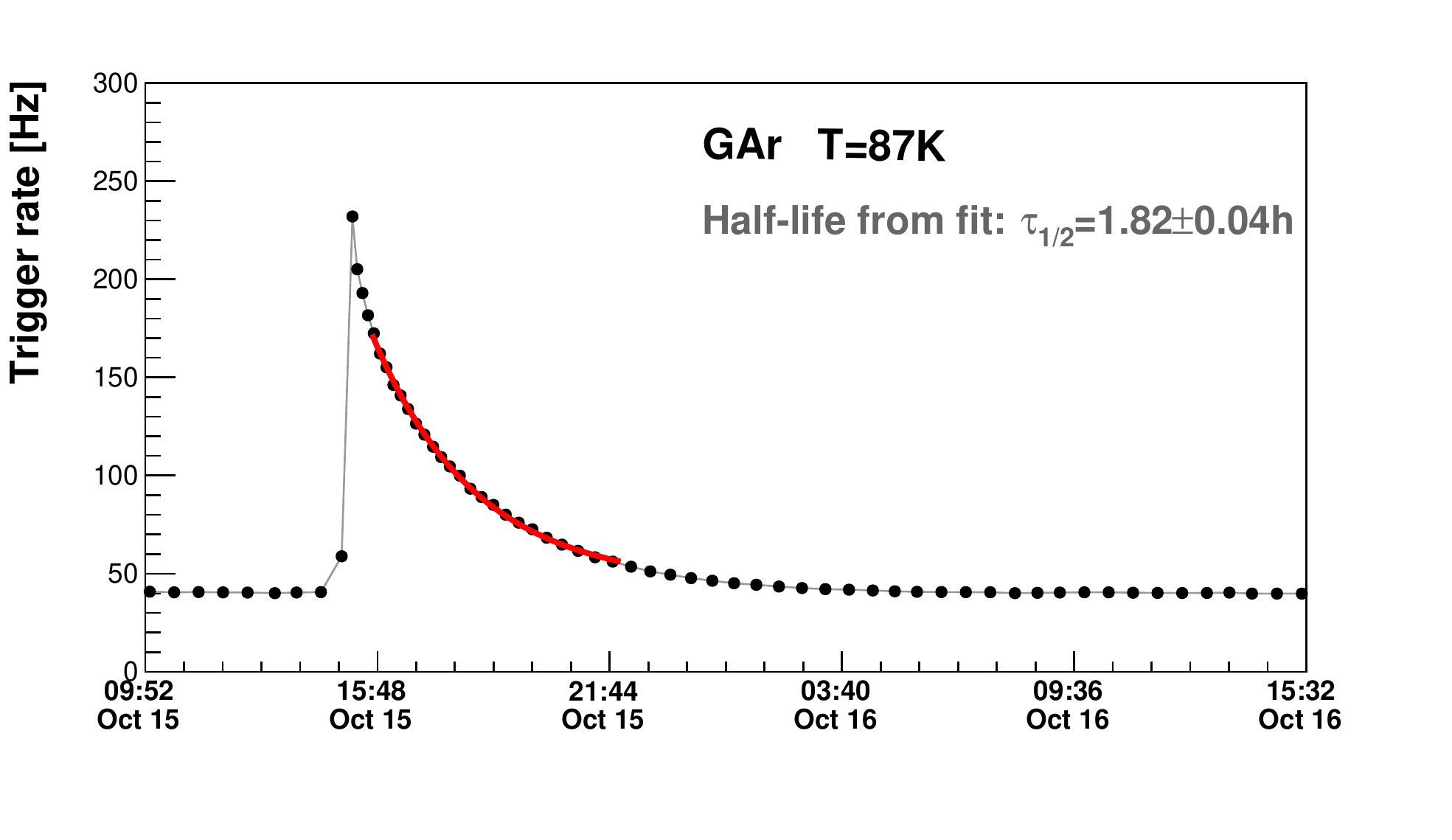}
\caption{Trigger rate (black dots) and fit (red line) in function of time around the injection of \kr\ atoms into the cold GAr target.}
\label{fig:krRate}
\end{figure}

A calibration campaign using the \kr\ calibration source was performed. The change in trigger rate after the injection of \kr\ is shown in Fig.\,\ref{fig:krRate}. The half-life of 1.82\,hours obtained from an exponential fit is in good agreement with the literature value of 1.83\,hours\,\cite{Kocher:1975}.

\begin{figure}[hbt]
\centering
\includegraphics[width=0.9\textwidth]{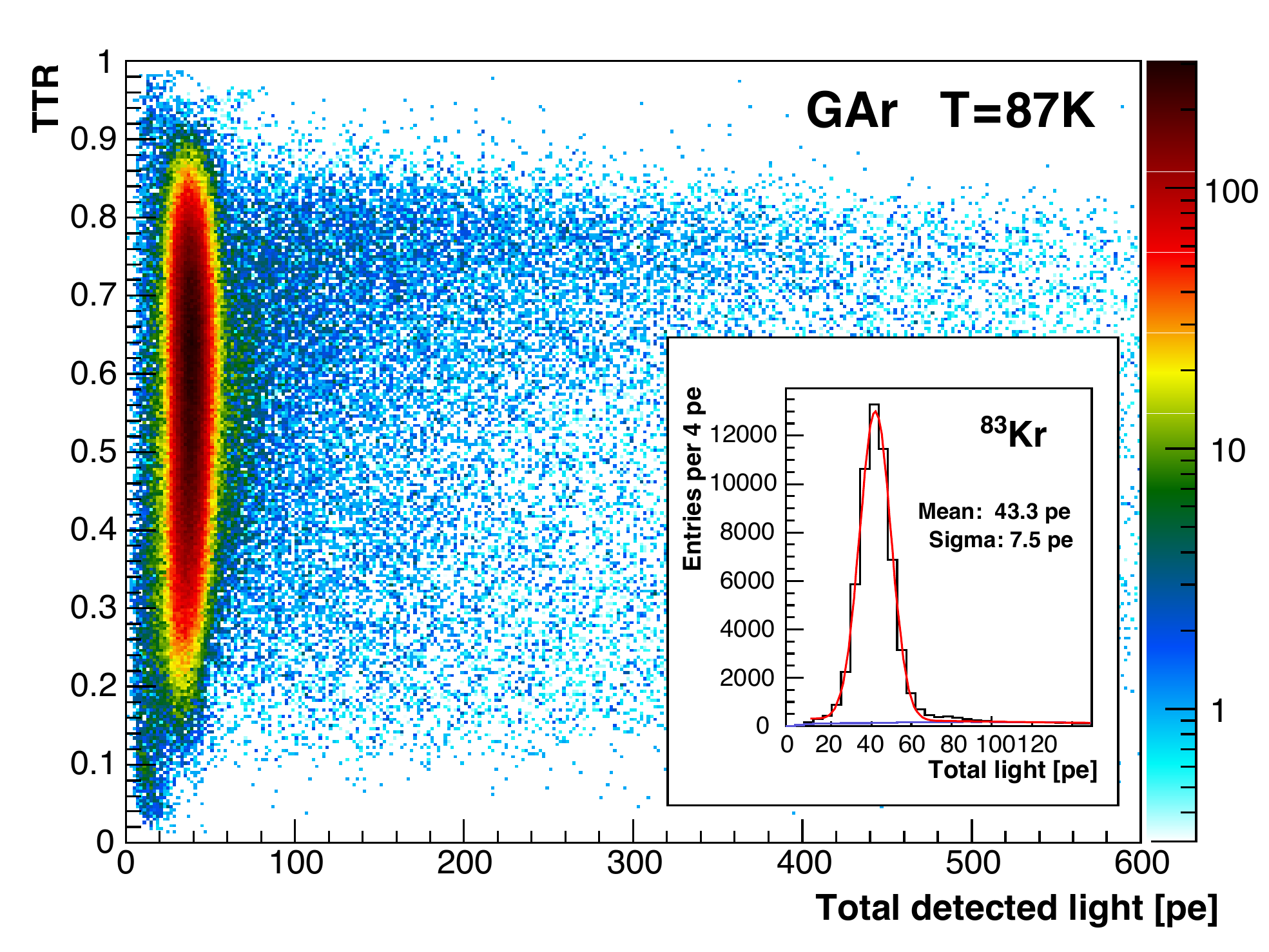}
\caption{$TTR$ vs. total detected light ($L_{\rm tot}$) of \kr\ events in cold GAr. The inset shows the projection of the data to the horizontal axis.}
\label{fig:Kr83mGArData}
\end{figure}

Figure\,\ref{fig:Kr83mGArData} shows a histogram of the vertical localization variable $TTR$ and the total detected light $L_{\rm tot}$ of 270k events obtained during the \kr\ injection. About 80\% of the data is related to \kr\ decays. The distribution of the events suggests a uniform light yield along the vertical axis and a spatially homogeneous distribution of the \kr\ isotopes. This conclusion is also supported by the Gaussian width of the spectrum (peaking at 41.5\kev ), which is close to the photo electron statistics, underpinning the functionality of the diffusive light collection. We note that the majority of triggers for these events are solely generated by the fast component ($\sim$30\%) of the scintillation light of the 32\kev\ transition of the \kr\ de-excitation. 

\subsection{Data with LAr target in single phase}
\label{sect:KrCalLAr}

Data taking with a full liquid argon target was performed for a duration of 6 months. The filling process itself required several weeks to cool and condensate in total almost 2 tons of argon gas (ALPHAGAZ\,2). The detector was readout at all times to continuously monitor its state. The filling level could be controlled by the capacitive level sensors, as well as by the increasing trigger rate due to the increasing target size. When the detector was full, a rate of 1.3\khz\ was obtained as expected from the natural abundance of \ar\ in the target ($\approx$1\,Bq/kg). The LAr purification system was operated continuously during the entire run and the  recirculation circuit was found to improve additionally the thermodynamic stability of the system. 
The slow scintillation decay time ($\tau_{\rm slow}$) was monitored continuously during the run and was found stable at a value around 1.25\mus . A direct measurement of the amount of oxygen in the LAr target by means of a trace analyser of the type AMI 2001RS\footnote{\url{http://www.amio2.com}} has been performed towards the end of the run. The concentration of oxygen impurities was measured to be less than 0.1\,ppm, at highest sensitivity of the instrument, suggesting the functionality of the purification system. More details can be found in \cite{Calvo:2016nwp}.

\begin{figure}[htb]
\includegraphics[width=0.9\textwidth]{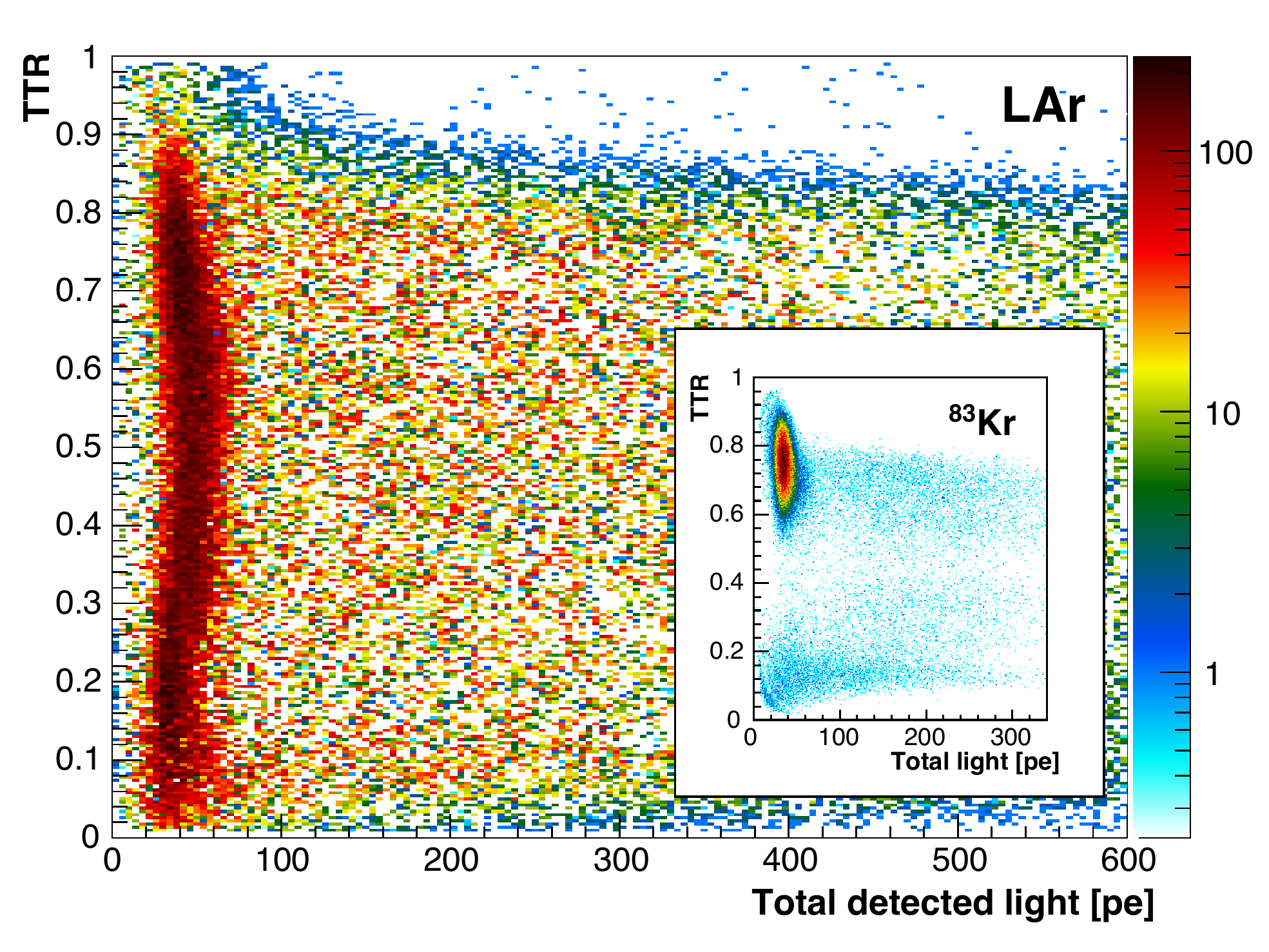}
\begin{center}
\caption{Distribution of \kr\ events in the LAr target after background subtraction (see text for explanation of the inset).}
\label{fig:Kr83mLArData}
\end{center}
\end{figure}

Several \kr\ calibration campaigns were undertaken during the run periode by injecting the metastable atoms in the gaseous phase on top of the liquid argon. The subsequent distribution of \kr\ atoms in the LAr volume can be derived from the $TTR$ values of events below the \kr\ peak. It was found that it took about 2 hours to obtain a spatially uniform distribution of the \kr\ atoms in the liquid. The long dilution time for the Kr atoms in LAr indicates a good performance of cryogenic design, suggesting an undisturbed, convection free condition of the LAr target. The main panel in Fig.\,\ref{fig:Kr83mLArData} shows \kr\ data taken at the full LAr target about 4 hours after injection. The background, dominated by \ar-\bet\ decays and external \gam\ photons, has been subtracted from the histogram. The inset shows the same data, but taken already 30--60\,min after injection, indicating the accumulation of \kr\ decays close to the liquid surface of the LAr target.

\begin{figure}[htb]
\centering
\includegraphics[width=0.9\textwidth]
{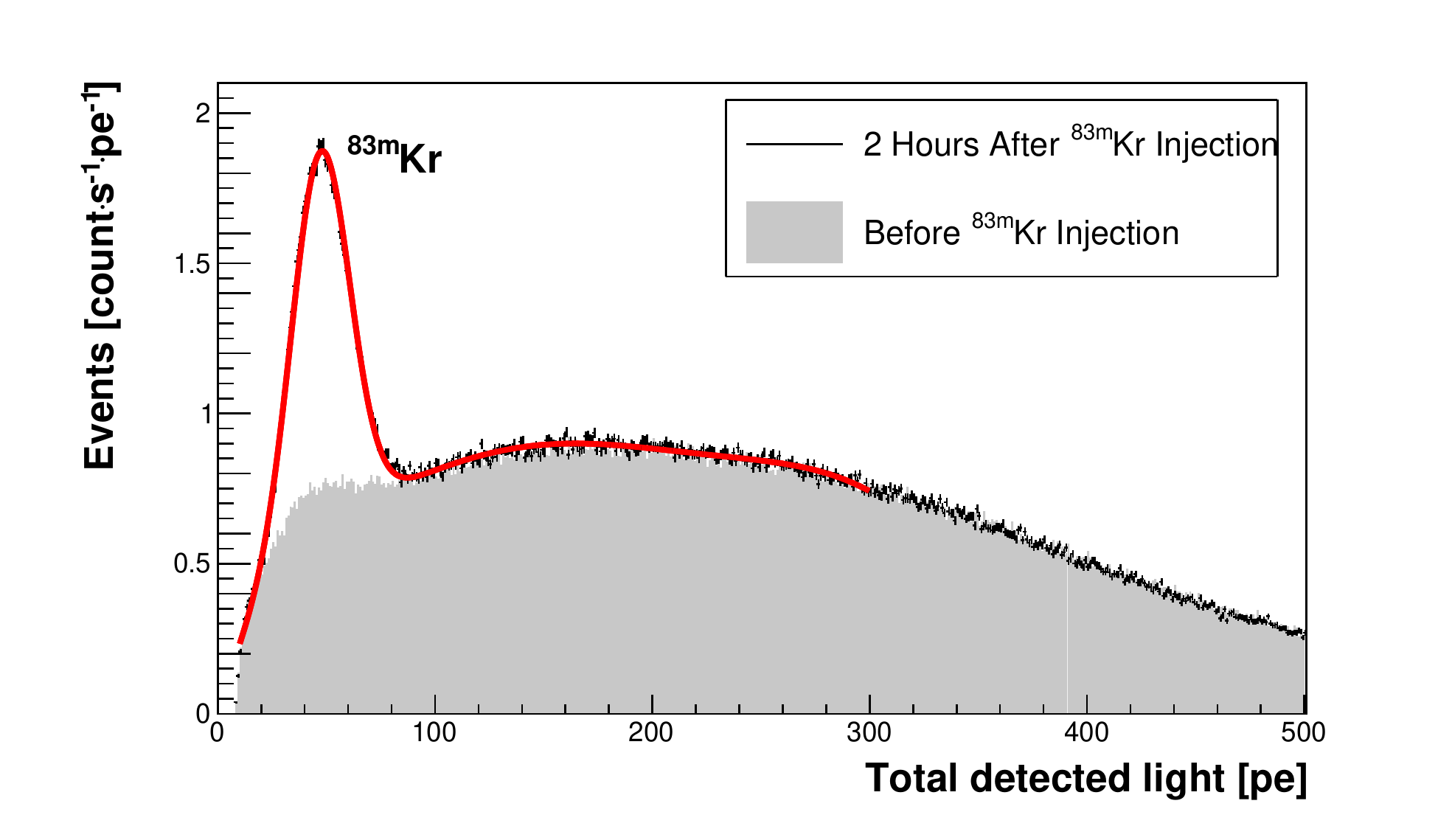}
\caption{Total light spectrum before (grey area) and 2\,h after (black dots) the injection of \kr\ into the LAr. The red curve is a fit to the data.}
\label{fig:krSpect}
\end{figure}

The total light spectrum from one of the \kr\ campaigns, obtained 2 hours after injection (black dots), is shown in Figure\,\ref{fig:krSpect} together with a fit function in red. The latter consists of a Gaussian on top of a background parametrisation by the sum of an 8th degree polynomial and an exponential. The background histogram before injection of the \kr\ atoms is shown in grey. Only events from 0.3\,$<$\,$TTR$\,$<$\,0.6 are used for this graph. 
The cleanly detected signal from the low energetic \kr\ events can be regarded as a proof of principle for recording and reconstruction of events at energies relevant for Dark Matter searches indicating the sensitivity of the ArDM detector to energies around $\sim$30\,keV$_{\rm ee}$.

The important experimental parameter of the light yield ($pe$/\kev) can be  estimated using the data from the \kr\ and \co\ campaigns. The Gaussian fit to the \kr\ peak yields a value of $LY_{Kr} = 1.1\,pe/\kev$. Similarly, a fit on the \co\ data gives $LY_{Co} = 1.2\,pe/\kev$.
The values obtained from both data sets agree well and we estimate a mean light yield of 1.1\,$pe$/\kev. Errors were found to be dominated by systematics, originating in fluctuations of temperatures, purities, calibrations, as well as the choice of fit ranges, background parameterisations and others. In total we estimate the systematic error on light yields to be around 5\%.  

The value for the light yield found in this work is lower than initially expected from Monte-Carlo simulations. In an extensive analysis of the measured light yield spectra and comparison to a description of the ArDM setup with a model of full light ray tracing, the attenuation length of the LAr target to its own scintillation light was determined to be of the order of 0.5\,m. This was explained by the presence of optically active impurities not filtered by the purification system (see Ref.\,\cite{Calvo:2016nwp} for more details). In absence of VUV attenuation within the argon bulk, the light yield is predicted to be $\simeq 2\,pe/$\kev.

\subsection{Detector stability}
\begin{figure}[hbt]
\begin{center}
\includegraphics[width=0.9\textwidth]{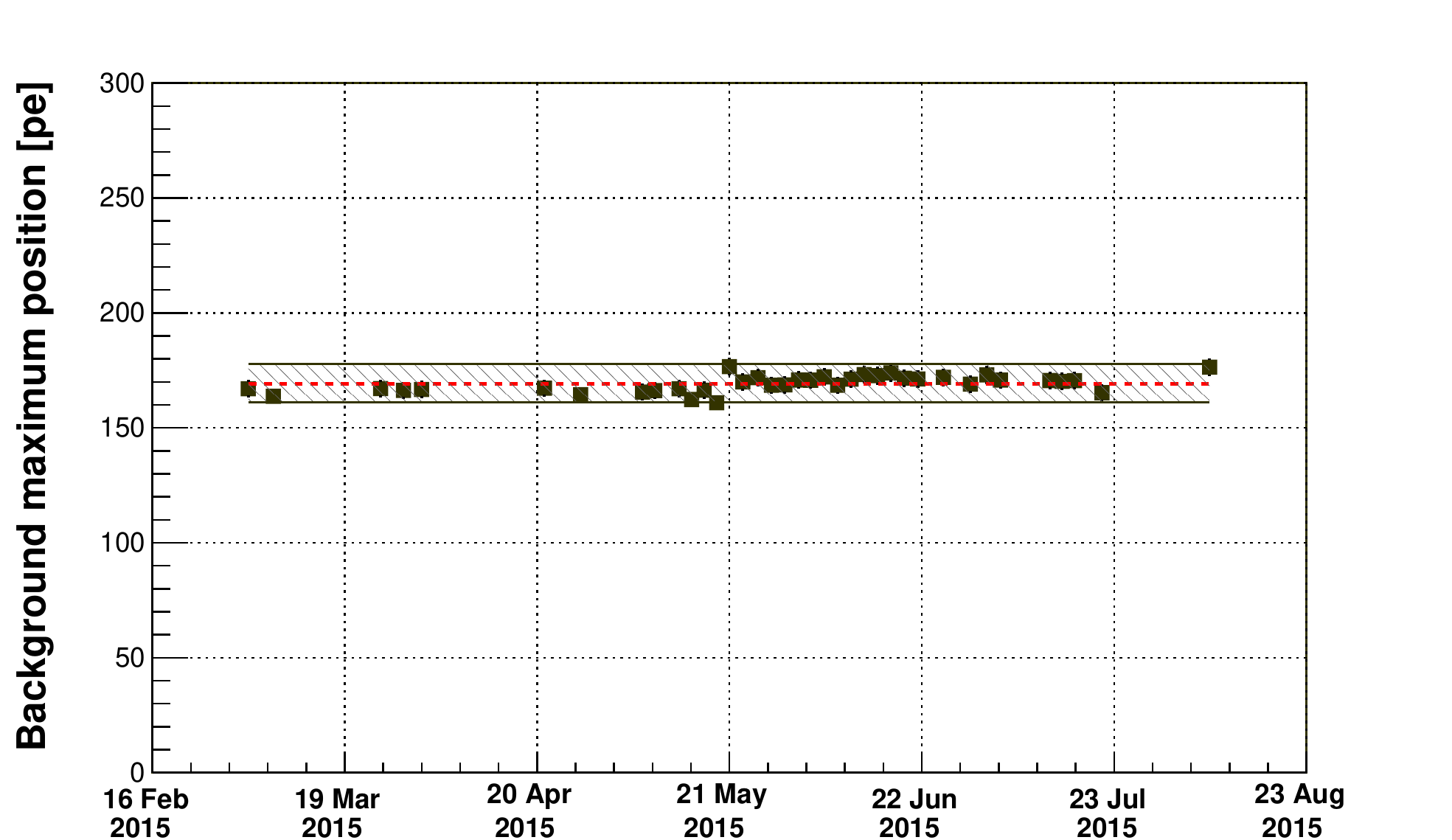}
\caption{The position of the maximum of the total detected light from background events as a function of time. The red dashed line shows the mean of the maximum position, 169\,$pe$, and the shaded area covers the $\pm$5\% uncertainty around the mean.}
\label{fig:ar39peak}
\end{center}
\end{figure}

One of the question addressed by the long-term operation of the detector was the durability of the thin evaporated layer of the TPB coatings on the inner surfaces, in total amounting to an area of about 3\,m$^2$. We analysed the variation of the position of the maximum of the total detected light and the light yield obtained from the \kr\ calibration campaigns as a function of time. We calculated the maximum value of the background from a polynomial function which parametrises the background. Figure\,\ref{fig:ar39peak} shows the time evolution of the background maximum position within 6 months, obtained from data randomly sampled from the recorded runs.
In addition, we compare the light yields ($pe$/\kev) in different \kr\ calibration campaigns. Table\,\ref{tab:kr_sum} shows a summary for the five \kr\ calibration campaigns during data taking with LAr. These measurements confirm the high level of stability of the ArDM setup.

\begin{table}[hbt]
\centering
\begin{tabular}{c r r }
\hline
Campaign  & \kr\,injection time  & Light Yield [$pe$/\kev] \\
\hline
1     & Apr. 20, 2015 & 1.07$\pm0.05$ \\
2     & Apr. 22, 2015 & 1.09$\pm0.05$ \\
3     & Jun. 02, 2015 & 1.07$\pm0.05$ \\
4     & Jun. 23, 2015 & 1.08$\pm0.05$ \\
5     & Jul. 17, 2015 & 1.10$\pm0.06$ \\
\hline
\end{tabular}
\caption{Light yield obtained from various \kr\ calibration campaigns. The errors shown are only due to systematic uncertainties.}
\label{tab:kr_sum}
\end{table}

In order to study the performance of the TPB wavelength shifter material during \ardm\ \rI, we compare the light yield ($pe$/\kev) from $^{222}$Rn and its progenies in room temperature GAr runs before and after data taking with LAr. From Table\,\ref{tab:rn_sum}, we confirm the good linearity of the light detection system before and after LAr filling. Within systematic errors, the measured light yield from both data set is consistent with each other.
 
\begin{table}[hbt]
\centering
\begin{tabular}{c c c c  }
\hline
  & Q-value & Before LAr Run  &  After LAr Run  \\
  & [\mev]  & Light Yield [$pe$/\kev] & Light Yield [$pe$/\kev]\\
\hline
$^{222}$Rn & 5.49  & 0.80$\pm$0.04 & 0.84$\pm$0.04 \\
$^{218}$Po & 6.00  & 0.81$\pm$0.04 & 0.85$\pm$0.04 \\
$^{214}$Po & 7.96  & 0.79$\pm$0.04 & 0.81$\pm$0.04 \\
\hline
\end{tabular}
\caption{The light yield from $^{222}$Rn and its progenies in room temperature GAr runs before and after detector LAr runs.}
\label{tab:rn_sum}
\end{table}
 
Both the background maximum position and the light yield from different \kr\ calibration campaigns show that the detector response was stable during data taking with LAr target. The good linearity and consistency of the light yield from $^{222}$Rn and its progenies before and after LAr runs shows the TPB wavelength shifter coated on the light readout system did not deteriorate during data taking with LAr target.
\section{Conclusion}

The \ardm\ detector is the first dual-phase liquid argon TPC optimised for Dark Matter searches to be operated in a deep underground environment. In this paper, we describe the current experimental setup configured for single phase mode. The cryogenic system, the light readout, the neutron shield, the DAQ and trigger as well as data reconstruction are detailed.

The detector was successfully commissioned underground at the Laboratorio Subterr\'aneo de Canfranc, first with warm gas and then cold gas. Then, it operated filled with liquid argon in stable conditions over six months. 
From a look at these first data, the performance and functionality are confirmed. 
The light yield shows good linearity in a wide energy range, from several tens of \kev\ to several \mev. Its absolute value is determined around 0.8\,$pe$/\kev\ in GAr and around 1\,$pe$/\kev\ in LAr.
The ability to detect, trigger and analyse signals below 30\,\kev relevant to Dark Matter searches is demonstrated. 

Following the successful operation reported in this paper, the detector has been recently upgraded for dual phase operation with a HV system, a new field cage and new reflectors. Furthermore, additional improvements e.g. in the purification system, are foreseen to improve the light yield.

The \ardm{} achievements represent an important milestone towards sensitive WIMP searches with liquid argon targets and open the path towards 10-tons or more scale detector with nuclear recoil sensitivity.

\section*{Acknowledgements}

We acknowledge the support of the Swiss National Science Foundation (SNF) and the ETH Zurich, as well as the Spanish Ministry of Economy and Competitiveness (MINECO) through the grants FPA2012-30811 and FPA2015-70657P, as well as from the "Unidad de Excelencia María de Maeztu: CIEMAT - FÍSICA DE PARTÍCULAS” through the grant MDM-2015-0509.

We thank the past and present directorates and the personnel of the Spanish underground laboratory {\it Laboratorio Subterr\'aneo de Canfranc} (LSC) for the support of the \ardm\ experiment. We also thank CERN for continued support of \ardm\ as the CERN Recognized RE18 project, where part of the R\&D and data analysis were conducted.

\bibliographystyle{JHEP}
\bibliography{ardmbib}{}

\end{document}